\documentclass[aps,prl,twocolumn,showpacs,amssymb,%
floatfix,superscriptaddress]{revtex4-1}
\usepackage[english]{babel}
\usepackage{amsmath}
\usepackage{verbatim}
\usepackage{amssymb}
\usepackage[usenames,dvipsnames]{xcolor}
\usepackage{epsfig}
\usepackage{soul}
\usepackage{natbib}
\usepackage[justification=justified, format=plain]{caption, subcaption}
\usepackage{epsf}
\usepackage{dcolumn}
\usepackage{bm}
\usepackage{hyperref}
\usepackage{url}
\usepackage{latexsym}

\definecolor{brightpink}{HTML}{F85bFa}
\definecolor{redpink}{HTML}{e91396}
\definecolor{greenyellow}{HTML}{9d9c07}
\definecolor{prussianblue}{HTML}{5771Fb}
\definecolor{lightblue}{HTML}{77beF3}
\definecolor{blackblue}{HTML}{2e3350}
\usepackage[english]{babel}
\usepackage{bm}
\usepackage{amsfonts}
\usepackage{amsthm}
\usepackage{graphicx}
\usepackage[latin1]{inputenc}
\usepackage{pgf}  

\hypersetup{colorlinks=true, citecolor=orange, urlcolor=cyan, linkcolor=blue, breaklinks=true}
\begin{document}
\title{Grand canonical validation of the bipartite International Trade Network}
\author{Mika J. Straka}
 \affiliation{IMT School for Advanced Studies, Piazza San Francesco 19, 55100 Lucca, Italy}
\author{Guido Caldarelli}
\affiliation{IMT School for Advanced Studies, Piazza San Francesco 19, 55100 Lucca, Italy}
\affiliation{Istituto dei Sistemi Complessi, CNR, Dip. Fisica Universit\`a ``Sapienza'', P.le A. Moro 2, 00185  Rome, Italy} 
\affiliation{London Institute for Mathematical Sciences, 35a South St, Mayfair London W1K 2XF, UK} 
\author{Fabio Saracco} 
\affiliation{IMT School for Advanced Studies, Piazza San Francesco 19, 55100 Lucca, Italy}

\begin{abstract}
Devising strategies for economic development in a globally competitive
landscape requires a solid and unbiased understanding of countries
technological advancements and similarities among export products. Both can be
addressed through the bipartite representation of the International Trade
Network. In the present paper, we apply the recently proposed
grand canonical projection algorithm to uncover country and product
communities. Contrary to past endeavors, our methodology, based on information theory, creates monopartite
projections in an unbiased and analytically tractable way. Single links between countries or products
represent statistically significant signals, which are not accounted for by
null-models such as the Bipartite Configuration Model. We find stable country
communities reflecting the socioeconomic distinction in developed, newly
industrialized, and developing countries. Furthermore, we observe product clusters based on the aforementioned country groups. Our analysis reveals the existence of a complicate structure in the bipartite International Trade Network: apart from the diversification of export baskets from the most basic to the most exclusive products, we observe a statistically significant signal of an export specialization mechanism towards more sophisticated products.
\end{abstract} 

\maketitle

\section{Introduction}
\noindent
The application of the network formalism in the field of socioeconomic science has seen an unprecedented growth in the last decades~\cite{barabasi2011takeover,caldarelli2010scale-free,pastorsatorras2010complex}.  Most of our actions take place in network environments and neglecting such structures can lead to insufficient interaction models~\cite{catanzaro2013network,jackson2014networks} and poor policy regulation~\cite{battiston2016complex}.
\newline\noindent
On the global scale, the analysis of the International Trade Network (ITN), also known as the World Trade Web, has taken a prominent role in the study of economic systems~\cite{serrano2003topology, Bhattacharya2008, barigozzi2010multinetwork, Mastrandrea2014a} motivated by the ongoing process of globalization. 
The ITN can be represented by a bipartite network in which the two layers are respectively countries and products~\cite{Hidalgo2009, Hausmann2011}: a link between the two layers is present if the selected country is able to export the chosen item. In this framework, \cite{Tacchella2012, Cristelli2013} proposed an algorithm that uncovers productive capabilities of countries, as well as the complexity of products.
\newline\noindent
Several works~\cite{Hidalgo2007, Caldarelli2012, Zaccaria2014} proposed different algorithms to infer relations among products from this bipartite network. However, very often such methods are either too tailored to the problem at hand and therefore lack generalizability, or the observed similarity between products neglects the validation by a statistical null-model so that the signal is indistinguishable from the  noise. 
\newline\noindent
Surprisingly, few general methods for projecting bipartite networks are present in literature, among which the seminal method proposed in~\cite{Tumminello2011}. Call respectively L and $\Gamma$  the two layers and suppose we want to project the bipartite network on the layer L; according to this method, the original bipartite network is divided into slices that are homogeneous in the degree of nodes of the opposite layer $\Gamma$. Comparing the presence of actual links with uniform distributions in each slice permits to establish the statistical significance of the observation and validate the co-occurrence of links between node pairs in the layer L. Even though this approach is poorly effective due to the high number of hypotheses to be tested (one for every couple of nodes in L for every slice), \cite{Tumminello2011} defines a controlled framework in which a statistical analysis can be performed. 
\newline\noindent
Recently, many efforts have been spent in providing an unbiased monopartite projection for bipartite networks~\cite{Gualdi2016, Dianati2016, Saracco2016}. Summarizing, through the comparison of the actual measurements with the expectations of the Bipartite Configuration Model (BiCM,~\cite{saracco2015randomizing}), it is possible to state if a node couple belonging to the same layer share a statistically significant fraction of their connections. 
In this process the BiCM may be too ``strict'' and could account for all the observations in the data~\cite{Saracco2016}: otherwise stated, the method may be too effective in reproducing original data. Although the original proposal embeds the BiCM~\cite{Gualdi2016, Dianati2016, Saracco2016}, the framework is very general and nothing prevents from using ``weaker" null-models~\cite{Saracco2016}.
\newline\noindent
In the present paper, we apply the aforementioned approach (discussed in details in~\cite{Saracco2016}) to the International Trade Network. The Configuration Model class is essentially obtained from entropy maximization and, since their derivation follows the general approach of \cite{Jaynes1957}, we will refer to this method as the \emph{grand canonical projection algorithm}.
\newline\noindent We observe that the BiCM induces a community structure which largely agrees with the socioeconomic distinction between developed, newly industrialized, developing and mainly raw material exporting countries.  Our analysis reveals a division within the group of developed countries around year 2000 into a core (Germany, USA, Japan, France, etc.) and a periphery (Austria, Italy, Spain, Eastern European countries, etc.), with the latter acting as a bridge to developing countries. 
\newline\noindent
The grand canonical projection shows also the presence of communities of products, which essentially reflect the development of their exporters. In particular, technological chemistry products cluster together because they are exported by the same developed countries, whereas electronic devices, textiles and garments form a community since they represent the typical exports of newly industrialized and developing countries. Each community of countries occupies the projected network of products in a particular way, focusing their efforts on few product communities, thus implying the presence of a statistically significant signal of specialization. 
Note that, usually, the picture arising from the analysis of the bipartite ITN is interpreted as the fact that the most developed countries export literally all possible products. In the present article, we refine this picture by highlighting that developed countries focus more on the most complex, i.e. technologically advanced, goods.
\newline\noindent A similar signal was already mentioned in the supplementary information of \cite{saracco2015randomizing}, though not discussed in details: the observed network appeared much more disassortative than the randomization, implying that countries with the largest export baskets link more than expected to products with the highest complexities, i.e. with the lowest degrees. These new results put in relation the topological network structure with economical meaning of the case study considered.\\
\newline\noindent
The paper is organized as follows. Firstly, we briefly introduce the ingredients of the grand canonical projection algorithm in the Methods section; more details on the BiCM and on the projection technique of the original article~\cite{Saracco2016} are provided in the Appendix A. The Results section presents the monopartite projections obtained from different null-models and their composition in terms of country communities, which reflect their stages in economic development, as well as product communities, based on their technological sophistication. Finally, we comment on the results and the performance of the methodology in the Conclusions.

\section{Methods}
\noindent 
We consider a binary bipartite network composed of two distinct node sets and a collection $E$ of undirected and unweighted edges. In the following, we distinguish the nodes sets by using Latin and Greek indices. The bipartite network is described by a binary biadjacency matrix $M$ of dimension $N_i \times N_{\alpha}$, where $N_i$ and $N_\alpha$ are the dimensions respectively of the Latin and the Greek layer and an edges between node couples $(i, \alpha)$ is represented by the matrix entry $M_{i\alpha} = 1$.\\ 
\newline\noindent 
In a bipartite network, the similarity between two nodes of the same layer is usually measured by the number of nearest neighbors in the oposite layer.
Even though this method is direct and intuitive, it neglects the crucial problem of determining which edges contain statistically relevant information and which do not. 
To address this question, in the present paper we employ the grand canonical projection methodology proposed in \cite{Saracco2016}, since it provides exact results and a coherent formalism. In the following, we will briefly review the method, inviting the interested reader to the original article.

\subsection{Grand canonical projection}
\noindent 
The grand canonical projection algorithm proposed in \cite{Saracco2016} yields a statistically validated monopartite projection of a bipartite network by comparing the observed node similarities with the expectations from a suitable null-model.  
\paragraph{Bipartite motifs as a measure of similarity} 
The number of common neighbors shared by two nodes of the same Latin (Greek) layer can be used as a measure of similarity. In the literature this quantity is known as the number of $K_{2,1}$ ($K_{1,2}$) bi-cliques~\cite{Diestel2017} or, following the formalism of~\cite{saracco2015randomizing}, as the number of V-motifs ($\Lambda$-motifs).  
\begin{figure}[h]
	\centering
	\includegraphics[scale=1]{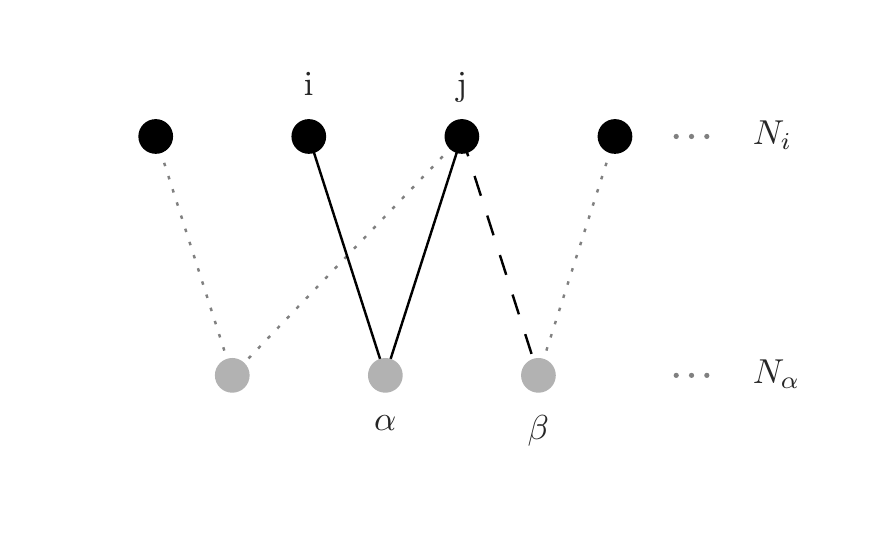}
	\caption{The $V^{ij}_{\alpha}$-motif is illustrated by the bold black edges between node couples $(i, \alpha)$ and $(j, \alpha)$.  Analogously, the edges between $(j, \alpha)$ and $(j, \beta)$ describe the $\Lambda_{\alpha\beta}^{j}$-motif. The top layer contains the ``Latin'' nodes, the bottom the ``Greek'' nodes. Other edges are indicated by dotted gray lines.} 
	\label{fig:lambda_motif}
\end{figure}
Fig.~\ref{fig:lambda_motif} illustrates the V-motifs between nodes i and j. The total number of V-motifs between these nodes is
\begin{equation}
	V^{ij}=\sum_\alpha V^{ij}_{\alpha} = \sum_\alpha M_{i\alpha} M_{j\alpha}.
\end{equation}
\paragraph{Bipartite Configuration Model and Bipartite Partial Configuration Models}
In order to establish whether two nodes are similar in a statistical sense, we compare the V-motif abundances with their expectations from the Bipartite Configuration Model (BiCM,~\cite{saracco2015randomizing}).  The BiCM is an entropy based null-model in which the information of the degree sequence of both layers is discounted. 
It has been shown~\cite{Hidalgo2007, Hidalgo2009, Hausmann2011, Caldarelli2012, Tacchella2012, Cristelli2013, Zaccaria2014,saracco2015randomizing} that the degree sequence is able to capture much of the structure of the bipartite international trade, as the triangular nested shape of the biadjacency matrix. Hence, discounting the degree sequence with a null-model allows to uncover more detailed substructures, which are otherwise hardly observable.
\newline\noindent The method is composed of two main steps: entropy
maximization and likelihood maximization. In the first step, we define
$\mathcal{G}_B$, an ensemble of bipartite graphs in which the number of links
can vary and the number of nodes per layer is fixed to the value of the network
under analysis. We define the Shannon entropy of the ensemble as
\begin{equation*}
S=-\sum_{G_B\in\mathcal{G}_B} P(G_B)\ln P(G_B),
\end{equation*}
where $P(G_B)$ is the probability of observing graph $G_B\in\mathcal{G}_B$.  
\newline\noindent
The choice of the Shannon entropy is standard in information theory and implicitly assumes that the system is ergodic, i.e. that all link configurations are theoretically possible. For non-ergodic or non-Markovian systems, the use of a non-additive entropy definition is recommended~\cite{Tsallis2004, Hanel2011}. In principle, different entropies could also be implemented in our case, although is has been shown that the Shannon entropy outperforms other entropy functionals in the context of behavioral networks~\cite{Squartini2015} in recovering missing information. 
%
\newline\noindent In the BiCM, we want to maximize the entropy constraining the degree sequence of both layers. Call $k_i$ and $k_\alpha$ the degrees of the node $i$ and $\alpha$, defined respectively on the Latin  and the Greek layer, and $\theta_i$ and $\theta_\alpha$  their relative Lagrangian multipliers. After the entropy maximization, the probability per graph reads
\begin{equation*}
	P\big(G_B |\theta_i, \theta_\alpha\big) =\prod_{i,\,\alpha}\dfrac{e^{-(\theta_i+\theta_\alpha)}}{1+e^{-(\theta_i+\theta_\alpha)}}
\end{equation*}
(see Appendix A for more details).
\newline\noindent The graph probability is a function of the unknown Lagrangian multipliers $\theta_i$ and $\theta_\alpha$. They can be estimated by maximizing the likelihood $\mathcal{L}=\ln P\big(G_B |\vec{\theta}\big)$ of observing the real network. It is possible to show that this approach is equivalent to imposing that the expectation values over the ensemble of the degree sequences are equal to those of the real network:
\begin{equation*}
\left\{
\begin{array}{c}
\langle k_i\rangle=\sum_\alpha\dfrac{e^{-(\theta_i+\theta_\alpha)}}{1+e^{-(\theta_i+\theta_\alpha)}}=k_i^*;\\
\\
\langle k_\alpha\rangle=\sum_i\dfrac{e^{-(\theta_i+\theta_\alpha)}}{1+e^{-(\theta_i+\theta_\alpha)}}=k_\alpha^*,\\
\end{array}
\right.
\end{equation*}
The values measured on the real network are marked with asterisks (*).
By solving the previous equations for $\theta_i$ and $\theta_\alpha$, it is possible to obtain the explicit value of the probability per graph in the ensemble.
\newline\noindent One of the most powerful properties of the BiCM is that it provides independent probabilities per link, which in turn permits to readily calculate the expectation values of more complicated quantities. For instance, the average number of V-motifs between $i$ and $j$ is
\begin{equation}\label{eq:expectedV}
	\langle V^{ij}\rangle^\text{BiCM}=\sum_\alpha p_{i\alpha}^\text{BiCM}\,p_{j\alpha}^\text{BiCM},
\end{equation} 
where $p_{i\alpha}^\text{BiCM}$ is the probability of observing a link between nodes $i$ and $\alpha$ according to the BiCM. 
\newline\noindent
Note that we could also impose non-linear constraints, such as the degree
variance of each node. However, this would lead to non-independent link
probabilities and complicate expressions such as eq.~(\ref{eq:expectedV})
significantly. Nevertheless, discounting the information of more elaborate
constraints, for example the bipartite clustering, may reveal other non-trivial
structures. Constraining the degree sequence thus represents a trade-off
between discounting non-trivial information and providing transparent and
easy-to-use tools for the analysis of bipartite networks.
%
\newline\noindent In the following, we compute the expected values of the
bipartite motifs presented in the previous paragraph according to the BiCM. For
the present objective, the constraints may turn out to be too strict, meaning
that they capture the main information contained in the network and no
statistical significant signal can be seen. In this case, it is opportune to
relax the constraints and fix the degrees of just a single layer. This model
has been proposed as the Bipartite Partial Configuration Model
(BiPCM,~\cite{Saracco2016}). Implicitly, the BiPCM$_\text{i}$, i.e. the one in
which only the degree sequence of the Latin layer is captured, is equivalent to
a BiCM in which all nodes in the Greek layer have degrees equal to their mean.
The BiPCM$_\text{i}$ is more effective in reproducing the observed number of
V-motifs rather than the number of $\Lambda$-motifs. Intuitively, the degrees
of the nodes $i$ and $j$ carry more information about $V_{ij}$ than the node
degrees of the opposite layer. Analogously, the BIPCM$_\alpha$ reproduces
$\Lambda$-motifs better than V-motifs.  
\newline\noindent 
In some cases, the projection of the real bipartite network can be completely reconstructed from its (bipartite) degree sequence, which means that the BiCM would be too strict to validate any links in the projection algorithm. The use of the BiPCM is thus recommended. By neglecting the information contained in the degree sequence of the layer opposite to the one of the projection, the BiPCM allows for stronger fluctuations stemming from the heterogeneity of the degrees which can be captured by the projection. A unique criterion for deciding a priori which null-model is more effective is currently missing as we are examining the limits of the different projections. This notwithstanding, as a rule of thumb we suggest that the BiPCM should be used when one deals with bipartite layers of very different lengths ($\frac{\text{longer layer}}{\text{shorter layer}} \gg1$) and one intends to project on the longer layer. Since the variability of the bipartite motifs is determined by the opposite layer, which is much shorter in this case, the BiCM is likely not to validate any links. In all other cases, the BiCM should be preferred.\\
\newline\noindent In the literature, the recent Curveball algorithm offers another way to discount the degree-sequence information in an unbiased null-model for bipartite networks.~\cite{Strona2014}. The authors implement a degree-sequence-preserving rewiring algorithm in order to build the ensemble of networks explicitly. Remarkably, the method is ergodic, i.e. it explores the phases space uniformly~\cite{Carstens2015} (note that the ergodicity of BiCM is automatically obtained by construction). Although the algorithm is relatively fast, the fact that it is micro canonical does not permit to calculate the expectation values of different quantities, thus preventing the possibility of writing an expression like eq.~(\ref{eq:expectedV}). In fact, $\langle V^{ij}\rangle^\text{Curveball}$ can be estimated as the average over a sample of the original ensemble defined by the Curveball algorithm. However, this sample has to be big enough in order to provide a sufficient statistics, i.e. to represent at best the whole ensemble without losing its statistical properties. For big networks, this procedure implies the presence of a large sample, which is hard to handle and increases the calculation times dramatically.
\paragraph{Statistical significance of node similarities: p-values and False
Discovery Rate} 
Both, the BiCM and the BiPCM, provide closed forms for the probability distributions of the $\Lambda$- and V-motifs. In the case of the BiCM, they follow a generalization of the binomial distribution, called Poisson-Binomial distribution \cite{Deheuvels1989, Volkova1996, Hong2013}, for each node couple. Depending on the constrained layer, the BiPCMs provide different distributions: for BiPCM$_\text{i}$, V-motifs follow a different Binomial distribution for each couple $(i,j)$, whereas the $\Lambda$-motif distribution is the same Poisson-Binomial distribution for every couple $(\alpha, \beta)$. The contrary happens for BiPCM$_\alpha$.\\
%
\newline\noindent 
By comparing the observed bipartite motifs with their null-model expectations, it is possible to calculate their associated \emph{p}-values, i.e. the cumulative probability of observing a value greater than or equal to the one actually observed. In a nutshell, the smaller the \emph{p}-value, the larger the similarity between the respective nodes compared to the null-model expectation. In order to validate the links between all the node couples in the same layer, a multiple hypotheses testing procedure should be adopted. In this framework, it is common to control the False Discovery Rate (FDR), i.e. the rate of falsely rejected null hypotheses~\cite{benjamini1995controlling}. Finally, the projection network can be obtained by drawing links whose \emph{p}-values is statistically significant according to the FDR test. In the following, the significance level of all validated networks is $\alpha=0.01$.\\ 
\newline\noindent 
We created an open-source implementation of the null-models and the calculation of the \emph{p}-values in Python, which are freely available on the web~\footnote{\url{https://github.com/tsakim/bicm}, \url{https://github.com/tsakim/bipcm}}. More details on the bipartite configuration models, the similarity measures and their distributions can be found in Appendix A.

\section{Data}
\noindent 
%
\newline\noindent
We use the BACI HS 2007 database from CEPII~\footnote{\url{http://www.cepii.fr/}} to construct the bipartite network, which comprises the export data for the years 1995 - 2010. Products are identified according to the Harmonized System and organized in hierarchical categories at different aggregation levels, which are captured by two, four, or six digit product codes. Here, we adopt the 2007 code revision (HS 2007) with four digit codes describing 1131 different products.
\newline\noindent 
In order to binarize data, it is customary to apply the  revealed comparative advantage (RCA), also referred to as Balassa index~\cite{balassa1965trade}, which describes whether a specific country is a relevant exporter of a product (RCA$\geq1$) or not (RCA$<1$) by comparing the relative monetary importance of the product in the country's export basket to the global average. The RCA is defined as
\begin{equation}
RCA_{c, p} = \left. \frac{e(c, p)}{\sum_{p'} e(c, p')} \middle/ \frac{\sum_{c'} e(c', p)}{\sum_{c', p'} e(c', p')} \right.,
\end{equation}
where $e(c, p)$ denotes the export value of product $p$ in country $c$'s export basket.

\subsection{Basic properties of the binary biadjacency matrix of the ITN} 
\noindent 
In the bipartite ITN, the degree distributions resemble a power-law for the countries and a Gaussian for the products. The degree heterogeneity can be approximately captured by the coefficient of variation (CV), i.e. the standard deviation over the mean, $\frac{\sigma}{\mu}$. As a rule of thumb, the larger the CV the less informative is the mean about the whole distribution. 
\newline\noindent The probabilities per link of the partial model BiPCM$_i$ (BiPCM$_\alpha$) are those of the BiCM in which the degree sequence of the opposite layer is approximated by its mean, i.e. $\langle k_\alpha\rangle=\frac{|E|}{N_\alpha},\,\forall\alpha$ ($\langle k_i\rangle=\frac{|E|}{N_i},\,\forall i$), where $|E|$ is the total number of edges. Since the CV varies between 0.5 and 0.55 for the products and between 0.82 and 0.89 for the countries, the BiPCM$_i$ will reproduce the V-motifs better between the countries than the BiPCM$_\alpha$ the $\Lambda-$motifs between the products. Generally speaking, the approximation implied by the partial null-models will work best for small CV and lose accuracy as the CV increases. 
%
\newline\noindent
In the trade data set we examine, the number of products is almost 10 times the number of countries and the biadjacency matrix is hence strongly rectangular. The connectance varies during the years between 0.09 and almost 0.13. This feature is related to the division of products in categories (see, for instance, \cite{Saracco2015}). 

\section{Results}
\noindent
The degree sequences of the binary bipartite trade network represent the sizes of country export baskets and the number of exporters of products, respectively. Implementing a null-model which discounts the information from the degree sequence (e.g. BiCM and BiPCM) implies focusing on structures that are not already contained in the heterogeneity of the degree distribution. For instance, in the BiCM the US export basket keeps its size while the composition of export products within the basket is randomized.  In the following, we shall observe effects in the structure of the international trade that are not explainable from dishomogeneities of the degree sequences alone.

\subsection{Country layer projection}
\noindent
The projection on the country layer induced by the BiCM reveals important information on different levels of economic development and the roles played by various countries in the globalization process.  
\begin{figure}[b!]
\begin{center}
\includegraphics[scale=.25]{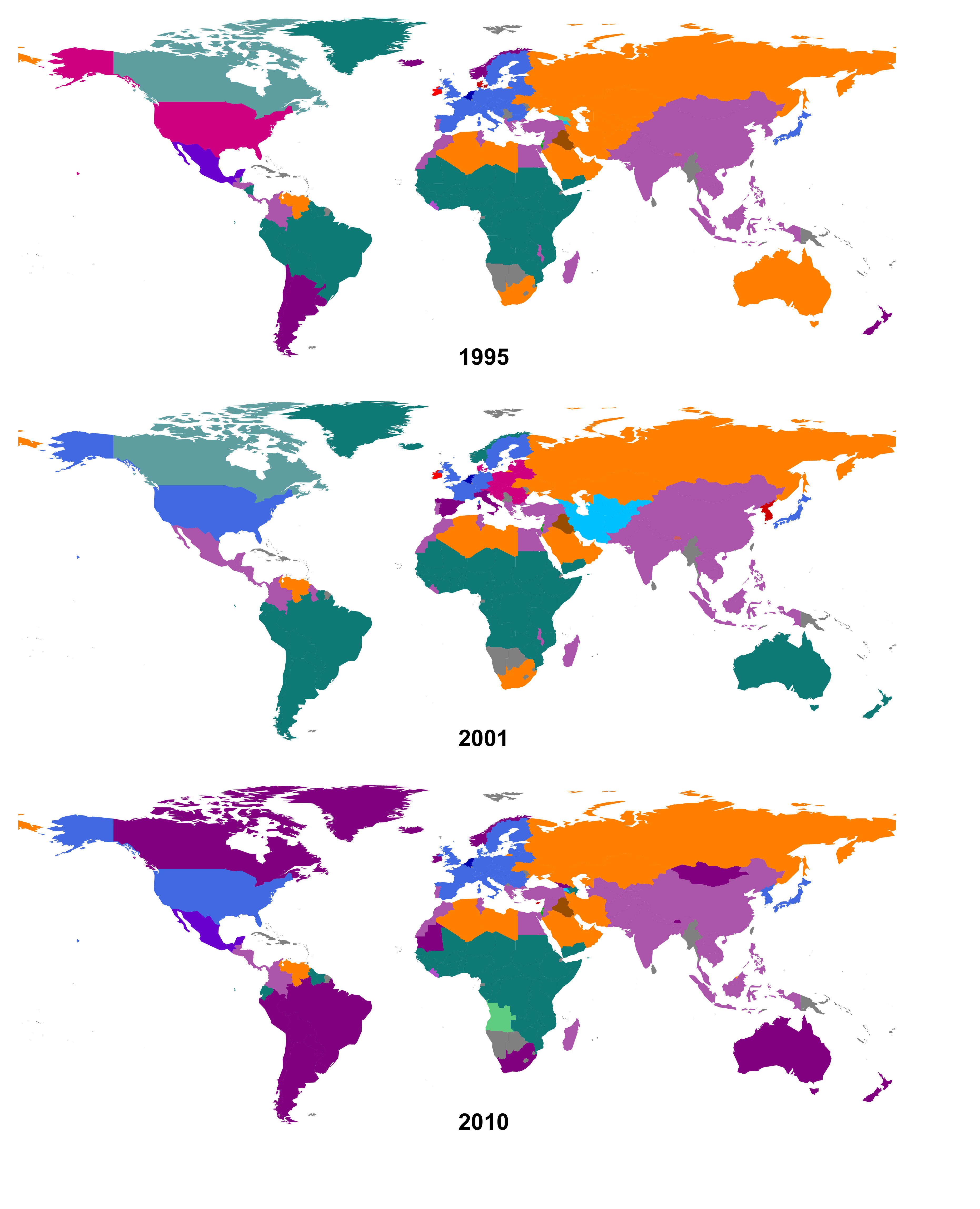}
\caption{Communities of countries based on BiCM projection for the years: 1995, 2001, 2010. Even though the division in communities show some noise, the partition in the following communities is stable: developed countries (blue/dark gray, see central Europe), newly industrialized and developing countries (light purple/lighter gray, see China), developing countries (green/darker gray, see central Africa), and countries whose exports rely on raw materials, e.g. oil (orange/light gray, see Russia).} 
\label{fig:BiCM_c_map_3}
\end{center}
\end{figure}
\begin{figure}[ht!]
\includegraphics[scale=.2]{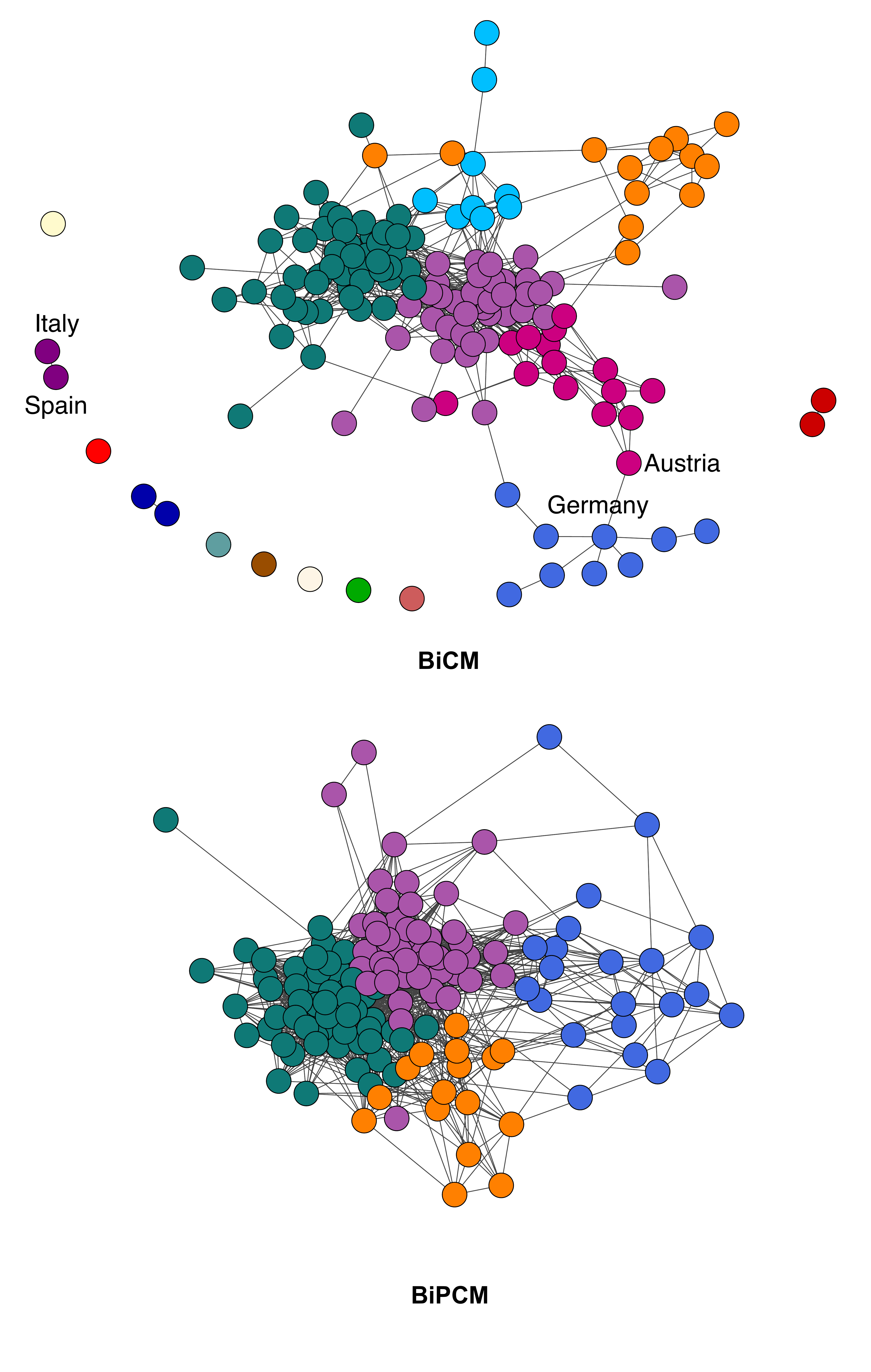}
\caption{Structure of the projected country network obtained with the BiCM and the BiPCM$_\text{c}$ for the year 2001. Note that in weakening the constraints, i.e. passing from BiCM to BiPCM$_\text{c}$, the connectance increases.}
\label{fig:BiCM_BiPCM_net}
\end{figure}
\newline\noindent 
As shown in Fig.~\ref{fig:BiCM_c_map_3}, an enhanced version of the Louvain community detection algorithm~\footnote{Since the Louvain algorithm depends on the order in which nodes are considered~\cite{Fortunato2010}, we implemented a more effective version of the original method. The community detection is repeated for several iterations with shuffled node sequence and the community partition giving rise to the highest modularity is eventually kept. The Python code can be found at \url{https://github.com/tsakim/Shuffled_Louvain}.} 
applied for the various years produces four stable clusters: developed countries (blue/dark gray),  newly industrialized countries (light purple/lighter gray),  African and South American developing countries (green/darker gray); developing countries exporting mainly raw materials such as oil (orange/light gray).
Despite some noise from year to year, mayor representatives of the blue community are Germany, USA, Japan, UK, and other European countries, while the purple community comprehends China, India, Turkey, Southeast Asia and some Central American countries; in the cluster of raw material exporters Russia, Saudi Arabia, Venezuela, post-Soviet states and North African countries can be found.
\newline\noindent Furthermore, we discern a fifth group whose composition fluctuates strongly during the considered time interval. It is mainly composed of countries with large coastal regions, which have little access to neighboring countries via continental trade routes. The community includes, among others, Australia, New Zealand, Canada, Chile, and Argentina. Much of their industrial output is aimed at internal markets and exports are strong in the fishing sector, especially for Canada and Chile. This explains why they are loosely linked to poorly industrialized nations like Mauritania, whose most important trade goods derive from fishing activities. As a result of the weak connectivity within the group, countries oscillate between different communities, which can clearly be seen, for example, for Australia and Canada in Fig.~\ref{fig:BiCM_c_map_3}. \\
\newline\noindent 
Relaxing the conditions of the null-model to just the degree sequence of the country layer leads to the BiPCM$_\text{c}$-induced projection.  The community structure is more stable than for the BiCM. In particular, note in Fig.~\ref{fig:BiPCM_c_map_3} that the fluctuating community disappears and the division of countries is more static. Weakening the constraints of the null-model thus reduces the noise in the projection. As a matter of fact, neglecting the constraints on the product layer means considering just the mean of the product degree sequence. The approximation is more accurate the smaller the relative dispersion of the product degrees, which is captured by the coefficient of variation and amounts to CV$\sim0.5$ in the present case (see ``Data'' section).\\ 
\begin{figure}[t!]
\includegraphics[scale=.25]{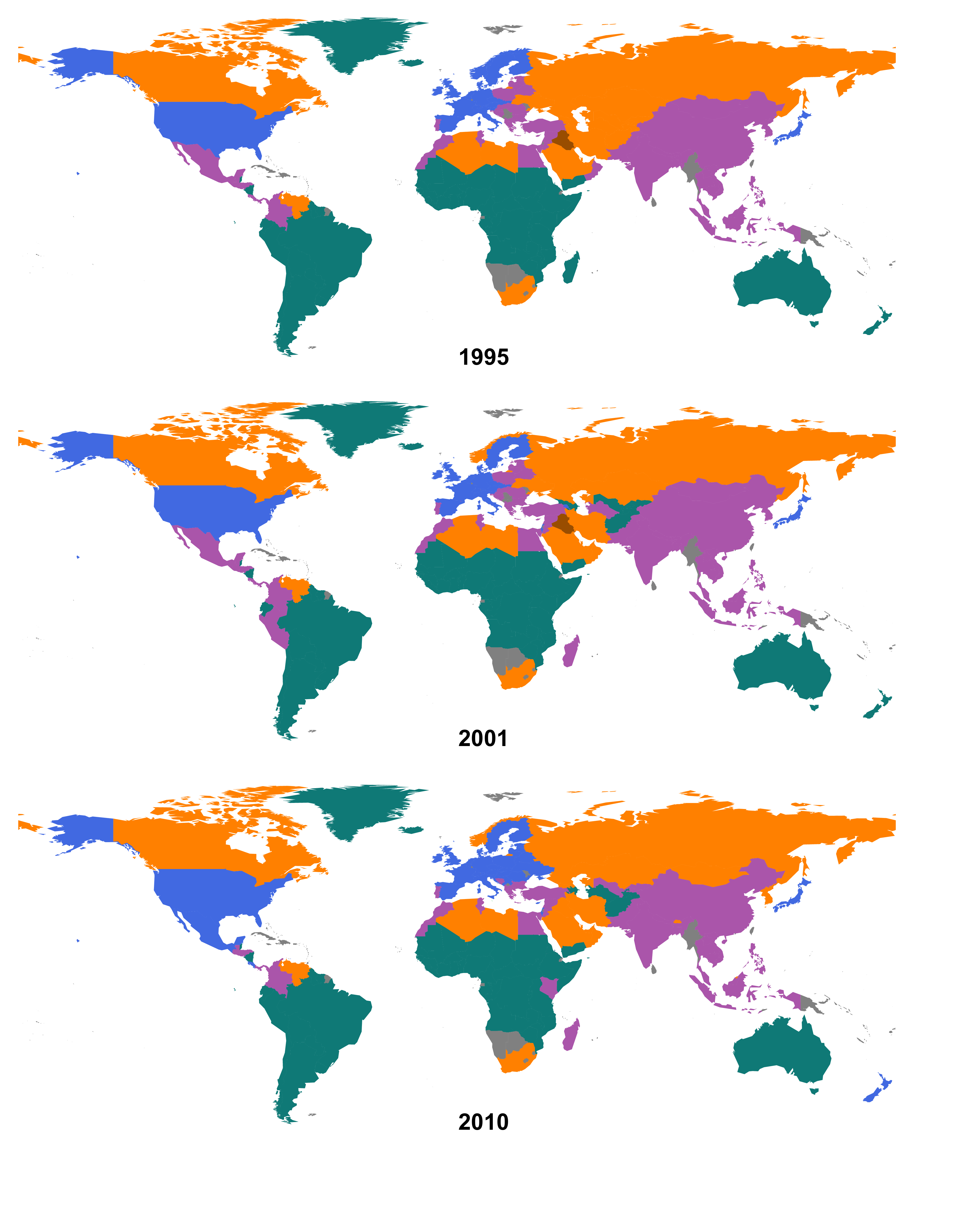}
\caption{Country communities based on the BiPCM$_c$ projection for the years: 1995, 2001, 2010. Compared to the BiCM communities of Fig.\ref{fig:BiCM_c_map_3}, the partition here is more stable.}
\label{fig:BiPCM_c_map_3}
\end{figure}
\newline\noindent
The downside of the stability of the BiPCM$_\text{c}$ projection is that it covers small, but insightful, structural changes. For instance, the BiCM manages to capture the split-off of Italy and Spain from the developed countries, as well as the separation of the developed European countries in an Eastern and a Western part during the years 1997-2002. As can be seen in Fig.~\ref{fig:BiCM_BiPCM_net}, Germany and Austria form a bridge between the Western and Eastern nations, with the latter themselves connecting to developing countries. \\
\newline\noindent 
Another striking result of the analysis of the country projection is the fact that many post-Soviet states still share a similar economic development years after the dissolution of the Soviet Union. A similar signal was detected in \cite{Saracco2016a}.

\subsection{Product layer projection}
\begin{figure}[t!]
\includegraphics[scale=.24]{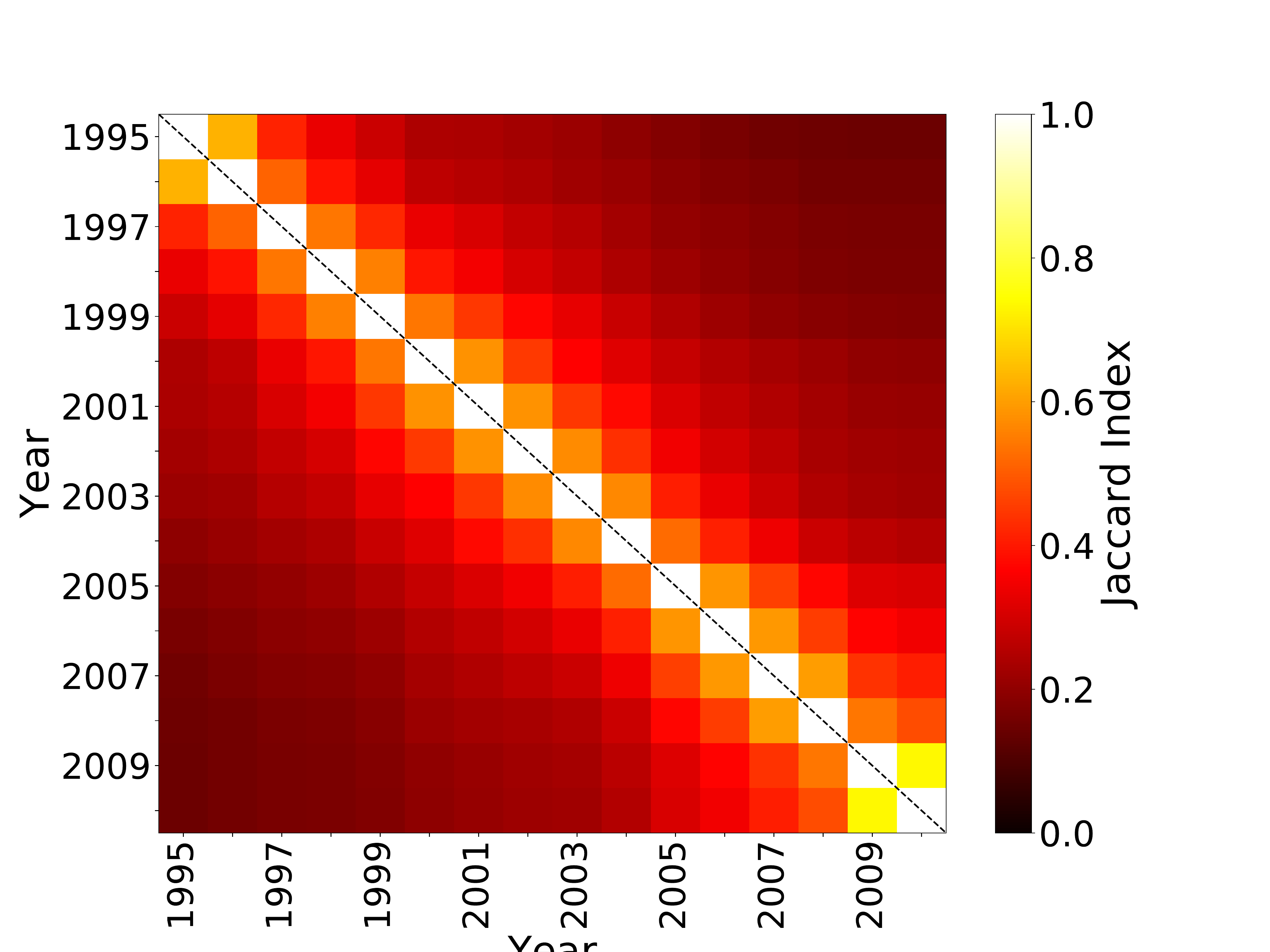}
\caption{Comparison of the product networks for the years 1995-2010. The
	Jaccard index measures the similarity between their edge sets, $E$, and
	is defined as {$|E_{year_i} \cap E_{year_j}|/|E_{year_i} \cup
	E_{year_j}|$}. The values fall very quickly below 0.5 for $|year_i -
	year_j| > 2$.
	} 
\label{fig:jaccard}
\end{figure}

\noindent
The BiCM-induced projection of the bipartite ITN network on the product layer does not reveal any statistically significant links, as already mentioned in~\cite{Saracco2016}. In other words, the total degree sequence of both countries and products contains enough information to account for the observed product similarities in terms of the $\Lambda$-motifs.
\newline\noindent
This observation stands in stark contrast to the country projection and is mainly due to two reasons connected to the different cardinalities of the layers. Firstly, the effective \emph{p}-value threshold for the validation procedure is proportional to the ratio of the significance level $\alpha$ over the number of tests that have to be performed, i.e. $\propto\alpha/\binom{N}{2}$ for $N$ nodes, as shown in Appendix A. Hence, the statistical validation is more restrictive on ``longer'' layers. In our case, the product layer is almost ten times larger than the country layer, which leads to a comparatively smaller effective threshold level.
\begin{figure*}
\begin{center}
	\includegraphics[scale=0.4]{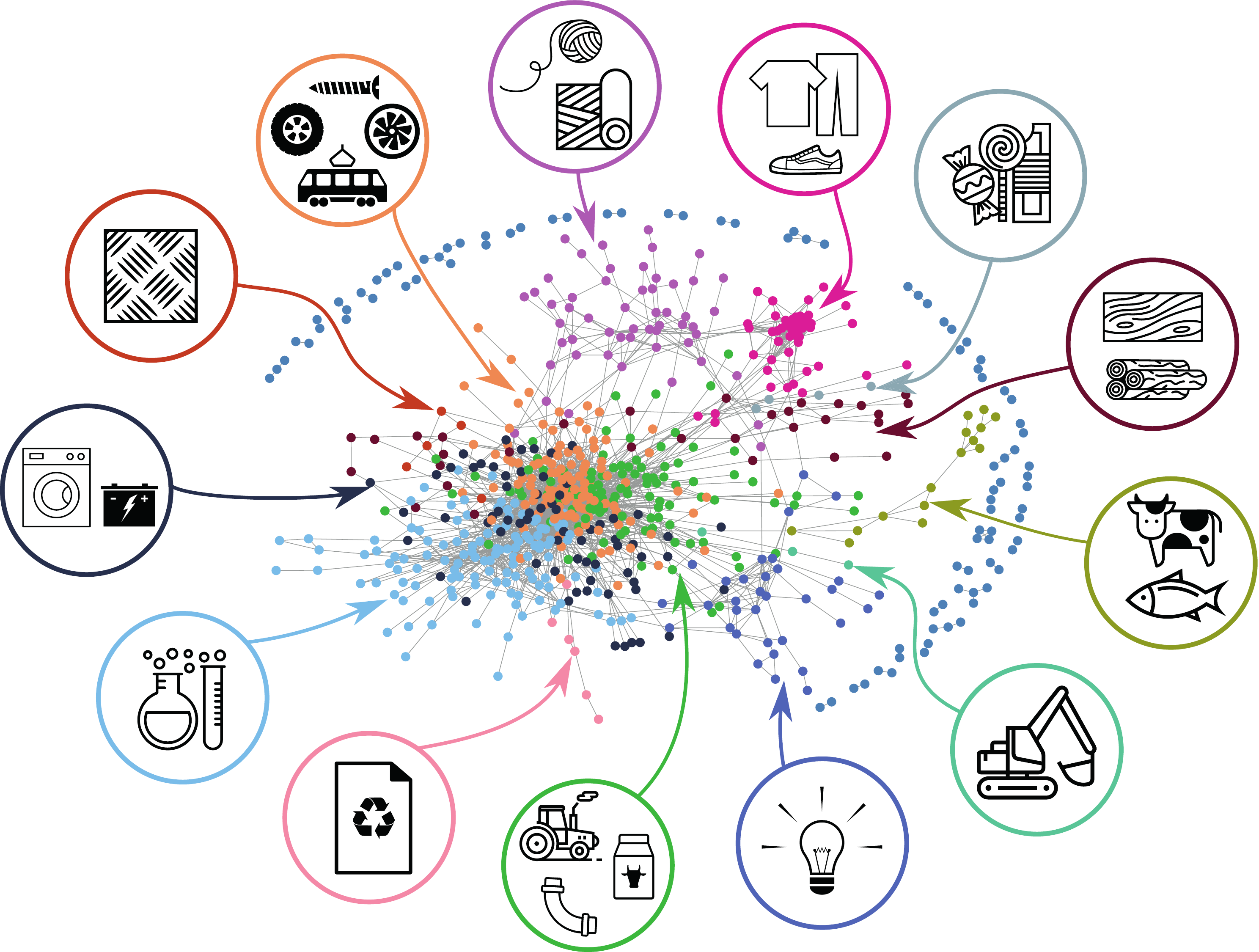}
	\caption{BiPCM$_\text{p}$ product network spanned by FDR validated
	edges for $\alpha = 10^{-2}$ in the year 2000. The communities have
	been obtained using the Louvain algorithm and include the following products, starting on the top and going clockwise:
	\textcolor{brightpink}{\textbullet} fabrics, yarn, etc.;
	\textcolor{redpink}{\textbullet} clothes, shoes, etc.;
	\textcolor{Brown}{\textbullet} wooden products;
	\textcolor{greenyellow}{\textbullet} animal products;
	\textcolor{prussianblue}{\textbullet} basic electronics;
	\textcolor{lightblue}{\textbullet} chemicals;
	\textcolor{blackblue}{\textbullet} machinery;
	\textcolor{Orange}{\textbullet} advanced electronics and machinery.}
	\label{fig:bicmi_fdr_nw_a001}
\end{center}
\end{figure*}
\noindent 
Secondly, the variability of node degrees depends on the length of the opposite layer, as mentioned in the Methods section, since the degree of each node stays in the interval between one and the dimension of the opposite layer. The degree heterogeneity of the longer layer is thus generally more limited than the one of the shorter layer, which reduces the set of possible values of the bipartite motifs between products in the present case.\\
\newline\noindent 
Due to the behavior of the BiCM, we implemented the BiPCM$_\text{p}$ to perform the validation procedure for product similarities.
As mentioned in the Methods section, constraining product degrees is more effective in reproducing the $\Lambda$-motif distribution than constraining country degrees. However, BiPCM$_\text{p}$ is going to be less effective in reproducing $\Lambda-$motifs than BiPCM$_\text{c}$ in reproducing V-motifs, since the coefficient of variation for the countries $\text{CV}\simeq0.8$ indicates a higher loss of information when approximating the country degree sequence by its mean.
%
\newline\noindent
The BiPCM$_\text{p}$-induced product networks are sparse with connectances in the range of 0.009-0.013 and highly fragmented for the years 1995-2010. As shown by the Jaccard indices of the edge sets in Fig.~\ref{fig:jaccard}, they are quite dissimilar from year to year. In the country networks on the contrary, the value never falls below 0.75 and 0.8 for the BiCM and BiPCM$_\text{c}$, respectively.  Nevertheless, the signal of product similarity persists: in fact, the enhanced Louvain community detection algorithm discovers a community structure that is stable throughout the years. The projection pinpoints evidently close relationships and captures broad communities, which remain constant, although the single links do not.\\
\newline\noindent 
Going into detail, the BiPCM$_\text{p}$ network consists of many small clusters surrounding the largest connected component (LCC), see Fig.~\ref{fig:bicmi_fdr_nw_a001}~\footnote{Icons: `Cow' by Nook Fulloption, `Fish' by Iconic, `Excavator' by Kokota, `Light bulb' by Hopkins, `Milk' by Artem Kovyazin, `Curved Pipe' by Oliviu Stoian, `Tractor' by Iconic, `Recycle' by Agus Purwanto, `Experiment' by Made by Made, `Accumulator' by Aleksandr Vector, `Washing Machine' by Tomas Knopp, `Metal' by Leif Michelsen, `Screw' by Creaticca Creative Agency, `Tram' by Gleb Khorunzhiy, `Turbine' by Leonardo Schneider, `Tire' by Rediffusion, `Ball Of Yarn' by Denis Sazhin, `Fabric' by Oliviu Stoian, `Shoe' by Giuditta Valentina Gentile, `Clothing' by Marvdrock, `Candies' by Creative Mania, `Wood Plank' by Cono Studio Milano, `Wood Logs' by Alice Noir from the Noun Project. All icons are under CC licence.} for the year 2000. Most of the isolated clusters are composed of vegetables, fruits, and their derivatives, such as lettuce and cabbage, soybeans and soybean oil, or fruit juice and jams. Other connections are less trivial: lead ores and zinc ores, for instance, are typically present in the same geological rock formations and appear as an isolated component in the network.
\begin{figure}[t!]
	\begin{tabular}{c}
		\begin{subfigure}[b]{0.5\textwidth}
			\includegraphics[scale=0.2]{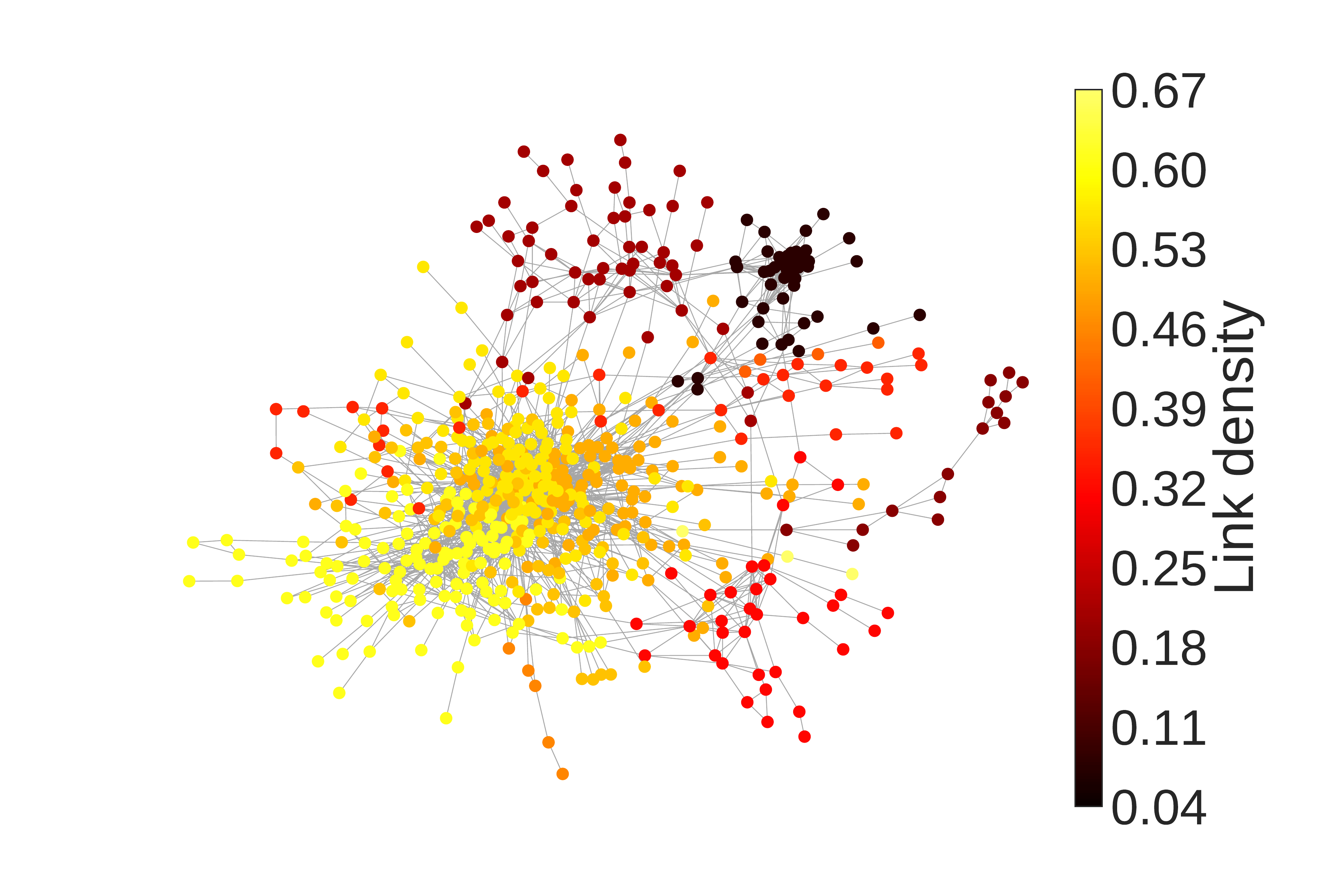}\\
		\end{subfigure}\\
		\begin{subfigure}[b]{0.5\textwidth}
			\includegraphics[scale=0.2]{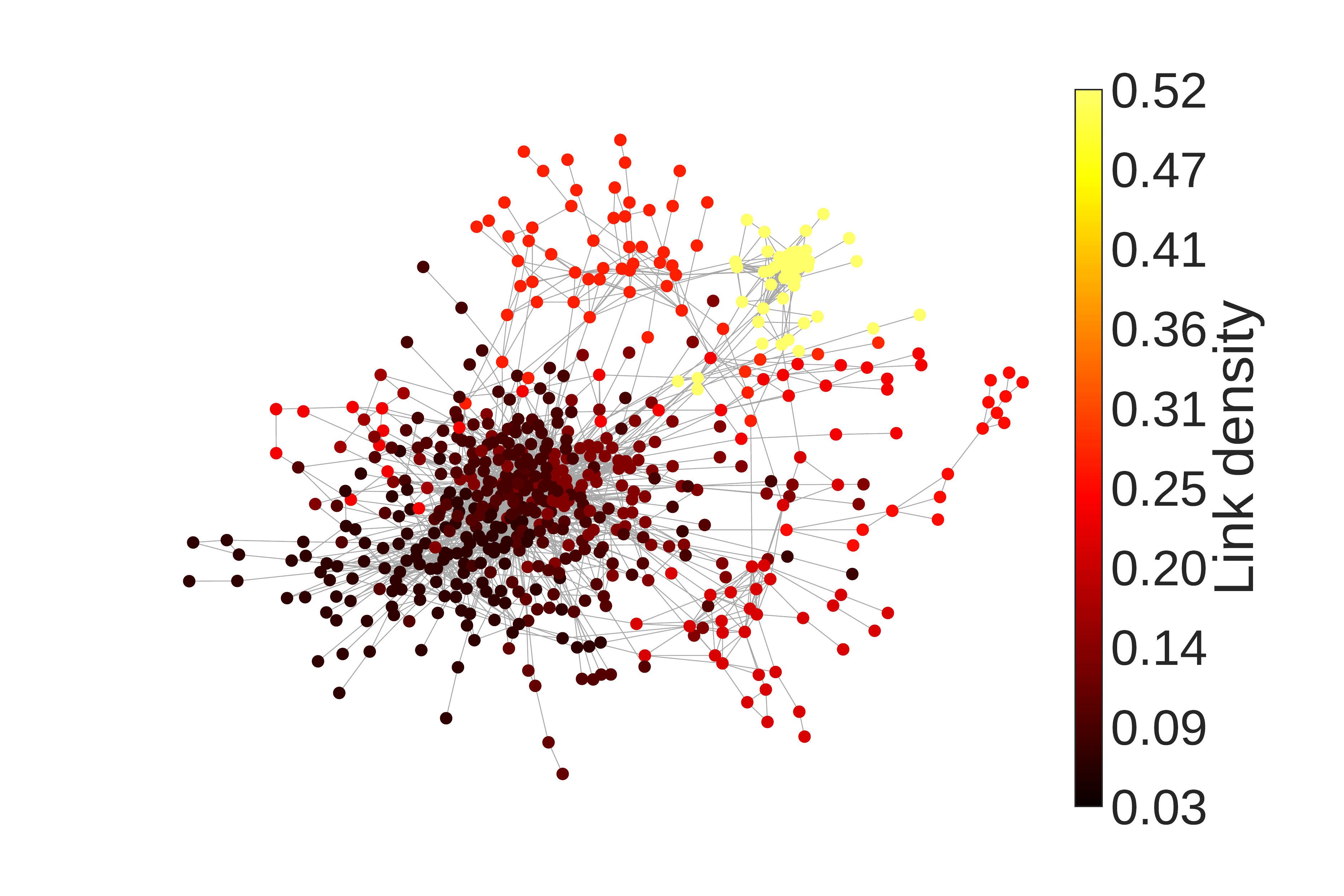}
		\end{subfigure}\\
		\begin{subfigure}[b]{0.5\textwidth}
			\includegraphics[scale=0.2]{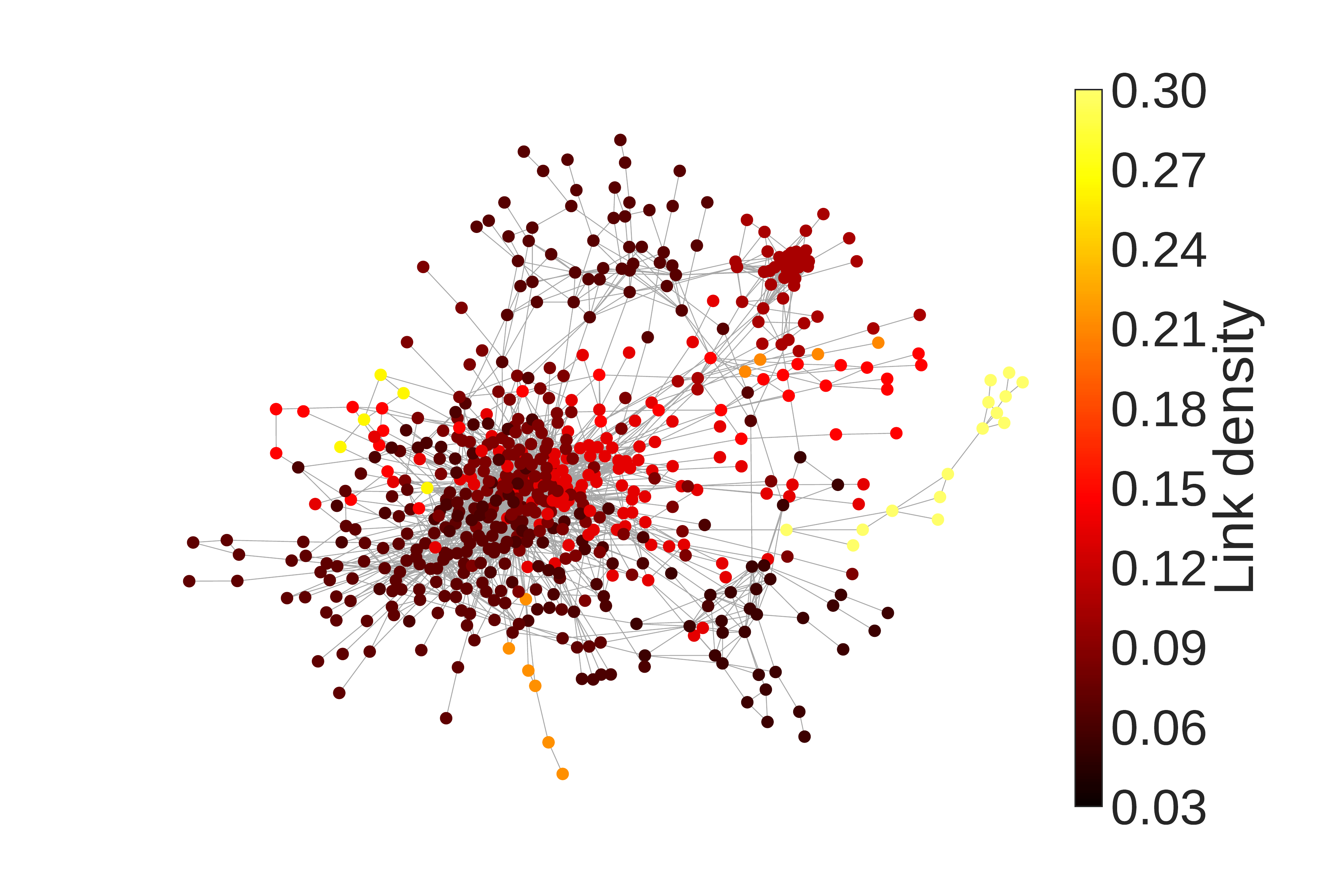}
		\end{subfigure}
	\end{tabular}
	\caption{The images show the relative focus of the country communities' exportation on different product cluster of the BiPCM$_p$ product network for the year 2000. Top: developed countries occupy the central communities of high technological and chemical products. Middle: developing countries focus on peripheral communities with relatively low complexity~\cite{Tacchella2012, Cristelli2013}. Bottom: raw material exporters are comparatively less focused, as shown in the link densities.} 
	\label{fig:Okk}
\end{figure}
\newline\noindent
The community detection algorithm uncovers a rich community structure inside the LCC, as shown in Fig.~\ref{fig:bicmi_fdr_nw_a001} for the year 2000. In the outer regions of the LCC we observe well-defined clusters, the most prominent of them being the garment and textile cluster that contains clothes and shoe products. Furthermore, one can discern a distinct community containing electrical equipment, such as circuits, diodes, telephones, and electrical instruments. Other clusters comprise bovine and fish products, yarns and fabrics, and goods made out of wood, such as planks, tool handles, etc.\\ 
\newline\noindent
The core of the LCC, on the other hand, hosts several overlapping communities containing mostly more sophisticated products, such as motors and generators, machines, cars, turbines, arms, chemical products, antibiotics, and other industrial products.  The community compositions are subject to fluctuations and include also, for example, agriculture and dairy products. The fuzziness of the core communities is due to the fact that they are typically exported by ``developed'' countries, which have large exportation baskets~\cite{Hidalgo2009, Hausmann2011, Tacchella2012, Cristelli2013}.\\
\newline\noindent
Note that the product communities do not follow necessarily the HS 2007 categorization, which is evident for the core communities where commodities of different origins can be found. As depicted in Fig.~\ref{fig:bicmi_fdr_nw_a001}, the green community, for example, is formed by milk, heavy-duty vehicles, and metal pipes. Although this may seem confusing at first sight, it is largely due to the fact that the projection derives originally from the exportation network and should reflect the different levels of industrialization of the exporting countries. This behavior is shown in Fig.~\ref{fig:Okk}: different country communities occupy mostly different product communities, as is captured by the index $I_{CP} = \frac{\sum_{i\in C, \alpha\in P} M_{i\alpha}}{|C||P|}$, i.e. the density of links between country community $C$ and product community $P$~\cite{Saracco2016}. Developed countries focus on the core communities and export, for instance, highly technological machinery and sophisticated chemical products.  At the same time, however, their export baskets encompass also products of low complexity such as milk and pipes, which are also exported, in fact, by newly industrialized countries next to textile products, garments, etc. In other words, the communities we observe, both on the product and the country layer, are derived from the way items interact: similar exports define countries with similar industrial development and, on the other hand, similar exporters define product communities of comparable technological level.\\ 
\newline\noindent The relative focus of country communities on specific product groups has strong implications. Evidence presented in studies on the bipartite representation of international trade \cite{Hidalgo2009, Hausmann2011,Tacchella2012, Cristelli2013, Hidalgo2007, Caldarelli2012, Zaccaria2014, Tu2016} connect productive capabilities to the triangular shape of the country-product biadjacency matrix, advocating that the most developed countries export even the least complex products. This stands in contrast to standard economic theories expressed by Ricardo~\cite{Ricardo1817}: according to his hypothesis, every country should specialize on the production of the most sophisticated goods its resources can support, even if they would be able to export less elaborate items as well.
\newline\noindent 
In past studies, the Configuration Models demonstrated their ability to uncover sub-structures and less evident information~\cite{Squartini2013u, Saracco2016a}. Already in~\cite{saracco2015randomizing} it was mentioned that the actual trade network is more disassortative than expected from the BiCM, implying that high degree countries (i.e. the ones with the largest export baskets) tend to export low degree products (i.e. the most exclusive and sophisticated ones) more than expected from the randomization.
\begin{figure*}
\begin{center}
\includegraphics[scale=0.3]{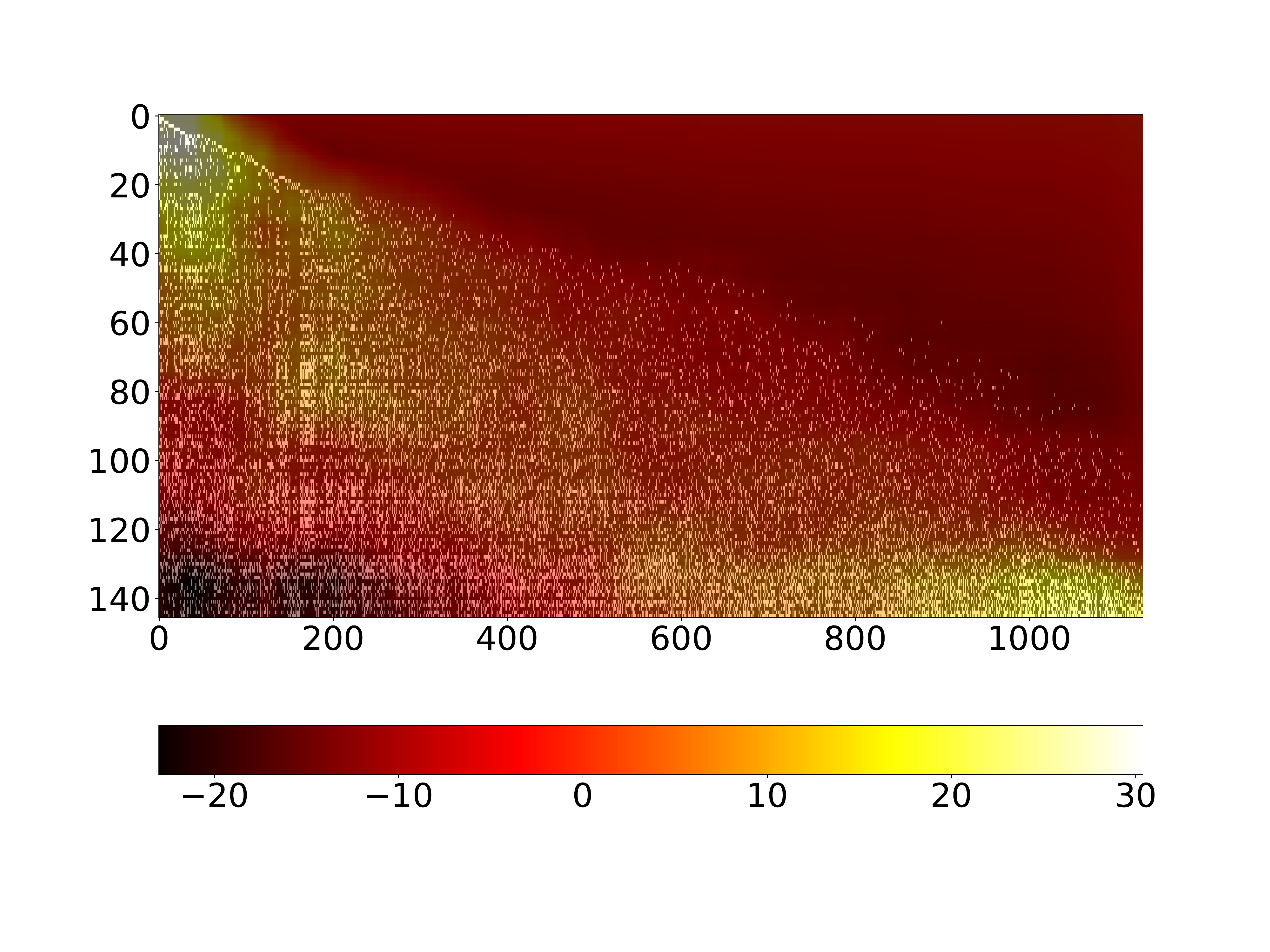}
\caption{ 
	Representation of the biadjacency matrix for the year 2000 with
countries along the rows and products along the columns, ordered by ascending
fitness and complexity ranking, respectively~\cite{Tacchella2012,
Cristelli2013}. Links are shown as white dots. The superimposed colors (gray
shading) correspond to the z-scores of the connectivity with respect to the
BiCM. The z-scores are calculated for boxes containing 20 countries and 80
products which are centered on the respective matrix entry. Lighter colors
indicate a higher presence of links than in the random null-model, darker
shades a lower one. As can be seen in the lower right corner, the most
developed countries (i.e. the bottom rows in the figure with the largest export
baskets) have higher densities that exceed the expectations from the null-model for
the most sophisticated products, i.e. those with the fewest exporters
($z\sim15$). On the other hand, the least developed countries with the smallest
export baskets focus their exports on basic products ($z\sim25$), as shown in
the upper left corner. In addition, the lower left part of the matrix shows
that high fitness countries export low complexity products much less than would
be expected from the BiCM. This indicates that countries export as many
products as they are capable of while focusing their efforts on the most
sophisticated commodities at the same time.}
\label{fig:biadj}
\end{center}
\end{figure*}
\newline\noindent 
Figure~\ref{fig:Okk}
explicitly shows that different countries, based on their technological level,
tend to focus of different areas of the product network. Otherwise stated, even
if the biadjacency matrix is triangular, still, once discounted the
contribution of the dimension of export baskets and the number of exporters, a statistically significant signal shows the presence of industrial specialization. In order to highlight this phenomenon, we compare the link densities in the
biadjacency matrix with the expectations from the BiCM null-model. For every
entry in the matrix, we consider a box of 20 countries $\times$ 80 products
that surrounds it~\footnote{Results are independent on the dimensions of the
boxes.}. We quantify the discrepancies between the observed number of links in
the boxes and their expectations from the BiCM using z-scores, i.e.
$z_\text{BiCM}(x)=\frac{x-\langle x
\rangle_\text{BiCM}}{\sigma_\text{BiCM}(x)}$. Z-scores express the difference
between the real value and the expectation in terms of the standard deviation:
$z\ll-3$ indicates that the observation is (significantly) less than the
null-model expectation, whereas $z\gg3$ is (significantly) more.  In
Fig.~\ref{fig:biadj} we represent the z-scores as a heat map on top of the
country-product biadjacency matrix. Links are shown as white dots. ``Hotter''
(lighter) areas are those where the actual number of links significantly
exceeds the BiCM expectation, whereas ``colder'' (darker) areas are those with
less links than expected. It is possible to observe two hot areas in the top
left and bottom right corner. The former shows that low fitness countries
export basic products much more than expected ($z\sim25$), whereas the latter
highlights the tendency of developed countries to export sophisticated products
($z\sim15$). Contrary to that, the bottom left corner illustrates that high
fitness countries export basic products much less than expected ($z\sim-20$).
It is possible to observe a ``hot'' area stretching from the top
left to the bottom right just below the diagonal of the matrix and a ``cold"
one just below that, highlighting the tendency of countries to focus on the
most sophisticated products they are able to export.


\section{Conclusions}
\noindent 
In the present paper we analyze the relations among countries and among products in the bipartite representation of the international trade network (ITN)~\cite{Hidalgo2009, Hausmann2011, Tacchella2012, Cristelli2013, Hidalgo2007, Caldarelli2012} by implementing a recently proposed strategy for the projection of bipartite networks~\cite{Gualdi2016, Dianati2016, Saracco2016}. The method is based on the Bipartite Configuration Model~\cite{saracco2015randomizing}, an entropy-based null-model discounting the information of node degrees. As a matter of fact, it has been shown that the degree sequence is responsible for the main characteristics of the trade network, such as the triangular structure of the biadjacency matrix between countries and products, see Fig.~\ref{fig:biadj}, \cite{Hidalgo2009, Hausmann2011, Tacchella2012, Cristelli2013, Hidalgo2007, Caldarelli2012, Zaccaria2014, Tu2016} Using the BiCM as a filter permits to uncover structures of the network not explained by node degrees. 
%
%
\newline\noindent 
The application of the BiCM to the ITN as a statistical null-model reveals communities of countries with similar economic development, namely developed, newly industrialized, and developing countries, and raw material (e.g. oil) exporters. These groups are stable throughout the years 1995-2010 except for some small deviations due to different progress in the ongoing globalization process. The communities become even more stable using the BiPCM (a weaker version of BiCM in which the degree sequence of only a single layer is constrained) for the monopartite projection. At the same time, however, the BiPCM is not able to detect smaller details like, for example, the post-Soviet state community, which is instead captured by the BiCM.
\newline\noindent 
Regarding the product layer, the BiCM turns out to be too restrictive to uncover any significant product similarities.  In other words, the information contained in the degree sequence of both layers is enough to account for the observed product relations in the data. Investigating the similarity among products therefore requires a relaxation of the constraints, logically leading to the application of the BiPCM. Such a phenomenon is essentially due to the rectangularity of the biadjacency matrix, i.e. to the dimension of the support of the distribution of bipartite motifs (for details see the Methods section). 
\newline\noindent
Using the BiPCM, we find product communities which define different industrialization levels and reflect the economic stages of their exporting countries. Highly sophisticated chemical products distinguish developed from newly industrialized and developing countries, whose exports focus mainly on electronic articles like diodes and telephones, or textiles and garments. It is worth pointing out that the communities are generally not due to productive chains, which should be reflected in a tree-like organization of the network. Observed clusters suggest that they are rather defined by the way countries organize their export baskets.
\newline\noindent Remarkably, our methods reveals a deeper structure than those discussed in \cite{Hidalgo2009, Hausmann2011,Tacchella2012, Cristelli2013, Hidalgo2007, Caldarelli2012, Zaccaria2014, Tu2016}. As already observed in previous studies, the biadjacency matrix of the country-product ITN is approximately triangular, which highlights the tendency of developed countries to export all possible products and not just the most exclusive ones. This observation conflicts with the Ricardo hypothesis, according to which countries should specialize their production on the most sophisticated products according to their resources. 
\newline\noindent However, as already mentioned, but not fully discussed, in the supplementary material of~\cite{saracco2015randomizing}, the real network appears more disassortative than expected by discounting the degree sequence. Otherwise stated, countries with a larger export basket tend to export more sophisticated products than expected. In the present paper we fully observe such a phenomenon through the different occupation patterns of product networks: different country communities with different technological levels tend to organize their export baskets differently, as shown in Fig.~\ref{fig:Okk}.
\newline\noindent The heat map in Fig.~\ref{fig:biadj} shows this phenomenon directly on the biadjacency matrix. The colors represent the z-scores encoding the discrepancy between the number of observed links in boxes drawn around the matrix entries and their expectations derived from the BiCM null-model (results are independent on the dimension of the box). Lighter colors represent abundances of links that are not explained by the null-model, whereas darker colors illustrate a lack in the links. Fig.~\ref{fig:biadj} shows that countries do not abandon the production of the most basic products, although they focus their exports on the most exclusive products. This can be seen by the ``hotter'' areas close to the diagonal, i.e. for the most exclusive products in the countries' export baskets. One can argue that the Ricardo hypothesis appears as a sort of second order effect: at first order the structure of the biadjacency matrix shows that the most developed countries are those with the largest export baskets (not those focused on most exclusive ones), at the second order a tendency to specialization is visible through a denser area for the most sophisticated products in the export basket.\\
\newline\noindent 
In summary, the grand canonical projection algorithm uncovers subtle structures
in the network under analysis: in the case of the world trade web, it reveals
an industrial specialization effect of country exports which is not appreciable
without the implementation of a null-model. This observation reconciles the
apparent contrast between recent studies that describe the development of
national productive capabilities in terms of the size of the export baskets on
the one hand, and standard economics and the Ricardo hypothesis expecting an
industrial specialization on increasingly complex products on the other hand.
From our analysis we can conclude that the degree sequence of the bipartite
network is responsible for the triangular shape of the country-product
biadjacency matrix, and thus for the former, whereas the specialization effect
is uncovered only once this information is discounted with the help of an
appropriately defined null-models. It is worth mentioning that both the
differentiation and specialization of countries are global and present
throughout the whole period of the analyzed data set. As shown
in Fig.~\ref{fig:BiCM_c_map_3}, local dynamics are observed through changes in
the community compositions depending on different local economic developments
and responses to global challenges. Nevertheless, the structure of the
International Trade Network as a whole remains constant over the years.\\
%
\newline\noindent We expect that the application of the grand canonical projection algorithm may reveal deeper structures even in other fields in which bipartite networks are heavily used. In biology, for example, statistically validated projections of mutualisitic network of pollinators and plants could uncover interaction patterns among pollinator species due to competition, for which measurements are rarely available and which remain generally unknown \cite{Bascompte2007,Suweis2013}.

\section*{Acknowledgments}
This work was supported by the EU projects CoeGSS (grant num. 676547), MULTIPLEX (grant num. 317532), Openmaker (grant num. 687941), SoBigData (grant num. 654024) and the FET projects SIMPOL (grant num. 610704), DOLFINS (grant num. 640772). The authors acknowledge Giulio Cimini, Riccardo Di Clemente, Tiziano Squartini and all participants to NEDO Journal Club @IMT for useful discussions.

\bibliographystyle{apsrev4-1}
%

\section{Appendix A: grand canonical Null-Models and Node Similarity}
\noindent
In the present appendix we revise briefly the methods of \cite{saracco2015randomizing, Saracco2016}, making use of the formalism introduced in the section Methods.

\paragraph{Bipartite Null-Models}
All configuration models of \cite{saracco2015randomizing, Saracco2016} are based on the statistical mechanics approach to complex networks~\cite{park2004statistical,Anand2009, Bianconi2013}; in this framework, the Shannon entropy per graph is defined as
\begin{equation}
	S = - \sum_{G_B \in \mathcal{G}_B} P\left(G_B\right) \ln P\left(G_B\right).
	\label{eq:shannon2}
\end{equation}
Here, $\mathcal{G}_B$ denotes the ensemble of bipartite graphs in which the number of nodes is constant, while the number of links can vary; $P$ is the probability of the bipartite graph $G_B$ belonging to ensemble $\mathcal{G}_B$. The entropy (\ref{eq:shannon2}) can be maximized subjected to the vector of constraints $\vec{C}(G_B)$. Solving the entropy maximization in terms of the probability per graph returns
\begin{equation}
	P\big(G_B |\vec{\theta}\big) =\frac{ \mathrm{e}^{-\mathcal{H}(\vec{C}(G_B),\vec{\theta})}}{Z\big(\vec{\theta}\big)},
\end{equation}
where $\mathcal{H}\big(\vec{C}(G_B),\vec{\theta}\big)$ is the Hamiltonian of the system, encoding the constraints. $Z(\vec{\theta})=\sum_{G_B \in \mathcal{G}_B} \mathrm{e}^{-\mathcal{H}( \vec{C}(G_B),\vec{\theta})}$ the partition function and $\vec{\theta}$ is the vector of Lagrangian multipliers associated to the entropy maximization.
\newline\noindent 
Different types of null-models can be obtained by modifying the constraints in the Hamiltonian.  For instance, by fixing the total number of edges $C= \sum_{i, \alpha} M_{i\alpha}=E$, we obtain the \emph{Bipartite Random Graph} (BiRG), a bipartite version of the well-known Erd\H{o}s-R\'enyi model~\cite{erdos1959random}. In accordance with the constraints, its Hamiltonian is given by
\begin{equation}
	\mathcal{H}_{BiRG} =\theta \cdot E.
\end{equation}
In this case both $\vec{C}$ and $\vec{\theta}$ are scalars, since there is just one condition. In the BiRG model, all edges are equally probable, with probability 
\begin{equation}\label{eq:pBiRG}
	p_{i\alpha}^\text{BiRG}=\frac{e^{-\theta}}{1+e^{-\theta}}\quad\forall i,\,\alpha.
\end{equation}
\newline\noindent 
In the so-called \emph{Bipartite Configuration Model} (BiCM) ~\cite{saracco2015randomizing}, the degrees of the nodes in both layers are constrained. If $k_i$ and $k_{\alpha}$ are the degrees respectively for the node $i$ in the Latin layer and for the node $\alpha$ in the Greek layer and $\theta_i$ and $\theta_{\alpha}$ the relative Lagrangian multipliers, the Hamiltonian reads thus
\begin{equation}
\mathcal{H}_{\text{BiCM}} = \sum_i\theta_i k_i + \sum_\alpha\theta_{\alpha} k_{\alpha},
\end{equation}
such that the probability is
\begin{equation}\label{eq:pBiCM}
	p_{i\alpha}^\text{BiCM}=\dfrac{e^{-(\theta_i+\theta_\alpha)}}{1+e^{-(\theta_i+\theta_\alpha)}}.
\end{equation}
\noindent
Relaxing the constraints of the BiCM yields the partial Bipartite Configuration Models, BiPCMs, introduced in~\cite{Saracco2016}. In particular, we constrain only the degrees of nodes in one layer. The corresponding Hamiltonians read
\begin{eqnarray}\label{eq:pBiPCM}
	p_{i\alpha}^{\text{BiPCM}_\text{i}} &=& \dfrac{e^{-\theta_i}}{1+e^{-\theta_i}}\quad\forall \alpha\\
	p_{i\alpha}^{\text{BiPCM}_{\alpha}} &=& \dfrac{e^{-\theta_\alpha}}{1+e^{-\theta_\alpha}}\quad\forall i.
\end{eqnarray}
It is worth pointing out that the Hamiltonians for the null-models defined above are all linear in the constraints. In fact the linearity of the constraints permits to express the graph probability $P\big(G_B)$ in terms of the single link probabilities, i.e. as
\begin{equation}\label{eq:probG}
	P(G_B) = \prod_{i, \alpha}^{N_i, N_{\alpha}} p_{i\alpha}^{m_{i\alpha}}\left(1 - p_{i\alpha}\right)^{1 - m_{i\alpha}},
\end{equation}
for any one of the null-model considered in the present section.\\
So far equations (\ref{eq:pBiRG}), (\ref{eq:pBiCM}) and (\ref{eq:pBiPCM}) are just formal, in the sense that the explicit values of the Lagrangian multipliers $\theta_i$ and $\theta_\alpha$ are unknown. In order to estimate them, following the strategy of \cite{Garlaschelli2008, squartini2011analytical}, we maximise the Likelihood of the ensemble on the real network. It can be shown that for the BiCM, it reads
\begin{equation}\label{eq:entmax}
\left\{
\begin{array}{c}
\langle k_i\rangle=\sum_\alpha\dfrac{e^{-(\theta_i+\theta_\alpha)}}{1+e^{-(\theta_i+\theta_\alpha)}}=k_i^*;\\
\\
\langle k_\alpha\rangle=\sum_i\dfrac{e^{-(\theta_i+\theta_\alpha)}}{1+e^{-(\theta_i+\theta_\alpha)}}=k_\alpha^*\\
\end{array}
\right.
\end{equation}
(starred quantities refers to the real network), such that the Likelihood maximisation is equivalent to impose that the degree sequence expectation values are equal to the values measured on the real network. The expressions for other null-models are analogous. Solving the system of equations (\ref{eq:entmax}) allows to calculate the values for all $\theta_i$ and $\theta_\alpha$ and explicitly obtain the probability per link.
\paragraph{Node Similarity}
In \cite{Saracco2016} the similarity measure implemented is just the number of bi-cliques $K_{2,1}$ or $K_{1,2}$ \cite{Diestel2017} (or V- and $\Lambda$-motifs, using the terminology of \cite{saracco2015randomizing}) existing between two nodes of the same layer. For instance, the number of all V-motifs between $i$ and $j$, $\text{V}^{ij}$, in the binary bipartite network is therefore given by 
\begin{equation}
	\text{V}^{ij} = \sum_{\alpha \in N_{\alpha}} M_{i\alpha}M_{j\alpha}.
\end{equation}
The ``standard'' approach is to consider the V-motifs as the quantity to analyse; in the approach of \cite{Saracco2016}, the statistical significance of every $\text{V}^{ij}$ is stated with reference to the aforementioned null-models in order to reveal relevant node similarities. The monopartite projection includes thus only edges $\left(i, j\right)$ whose relative $\text{V}^{ij}$ are statistically significant. Since edges are independent (see (\ref{eq:probG})), the probability of measuring a V-motif consisting of $(i,j)$ on the Latin layer and $\alpha$ on the Greek layer is
\begin{equation}
	P(\text{V}^{ij}_{\alpha}) = p_{i\alpha} p_{j\alpha}.
\end{equation}
In the case of the Random Graph model, for instance, $P(\text{V}^{ij}_{\alpha})_\text{BiRG}\equiv (p^\text{BiRG})^2\, \forall i,j \in N_i, \forall\alpha \in N_{\alpha}$, since the edge probability is independent of the couple $(i, \alpha)$ and uniform in the network. In this sense, the probability distribution of $\text{V}^{ij}=\sum_\alpha\text{V}^{ij}_{\alpha}$ is the sum of independent Bernoulli events, all with the same probability $(p^\text{BiRG})^2$, i.e. a Binomial distribution. In the Configuration Model, on the other hand, $p_{i\alpha}$ differs from couple to couple: $\text{V}^{ij}$ is thus the sum of independent Bernoulli random variables, in general with different success probabilities. The probability of observing $P(\text{V}^{ij} = k)$ will therefore be given by
\begin{equation}
	P(\text{V}^{ij} = k)=\sum_{\tilde{\alpha}_k\in A_k}\prod_{\alpha\in\tilde{\alpha}_k} P(\text{V}^{ij}_{\alpha})\prod_{\alpha'\notin\tilde{\alpha}_k} (1-P(\text{V}^{ij}_{\alpha'}) ) 
\end{equation}
where $A_k$ is the set of all possible choices of $k$ elements from the set $\{1,2,\dots, N_\alpha\}$ and $\tilde{\alpha}_k$ is a single realization,~\cite{Hong2013}. 
\newline\noindent 
The Partial Configuration Models are between BiCM and BiRG: the distribution for $\text{V}^{ij}$ is the same Poisson Binomial for all couples $(i,j)$ in the case of BiPCM$_\alpha$, while it is a Binomial distribution with probability $p=\frac{k^ik^j}{N_\alpha^2}$ for BiPCM$_i$. In fact, in reconstructing the structure of the V-motifs network it is much more effective to know the degrees of the nodes involved in the V-motifs than the nodes on the other layer, such that the BiPCM$_i$ resulted as more accurate in the network reconstruction. The limits of this intuition are currently under analysis.\\
\paragraph{Statistical Hypothesis Testing and FDR}
Validating the statistical significance of the measured $\text{V}^{ij}$ therefore revolves around the null hypothesis that its observed value can be explained by the underlying null-model, i.e. that it is compatible with the corresponding probability distribution. For this purpose, we calculate the \emph{p}-values for right-tailed tests, i.e. $P(\text{V}^{ij} \geq \text{V}^{*\,ij})$, where $\text{V}^{*\,ij}$ is the measure on the real network. Note that the total number of distinct couples $\left(i, j\right)$ and therefore the number of different hypotheses which are tested simultaneously is $N_i (N_i -1)/2$. Among other proposals for multiple hypotheses testing, the false discovery rate (FDR)~\cite{benjamini1995controlling} permits a control at each step of the verification procedure. The \emph{p}-values are ordered according to their values from smallest to largest and label by $k$. The largest $\hat{k}$ that satisfies
\begin{equation}
p_{value}^{\hat{k}} \leq \frac{\hat{k} \alpha}{N_i(N_i - 1)/2},
\end{equation}
defines the effective threshold $p_{th}=\frac{\hat{k} \alpha}{N_i(N_i - 1)/2}$: the hypotheses associated to all \emph{p}-values smaller than or equal to $p_{th}$ are rejected and are declared as ``statistically significant'', i.e. they cannot be explained by the null-model. Once the \emph{p}-value associated with the couple $(i,j)$ is rejected, in the projected network a binary link is drawn between the two nodes.

\section{Appendix B: Limits of the BiRG and the BiPCM$_\text{c}$ projection on the products layer}
\begin{figure}[h!]
\begin{tabular}{c}
	\begin{subfigure}[b]{0.5\textwidth}
		\includegraphics[scale=0.26]{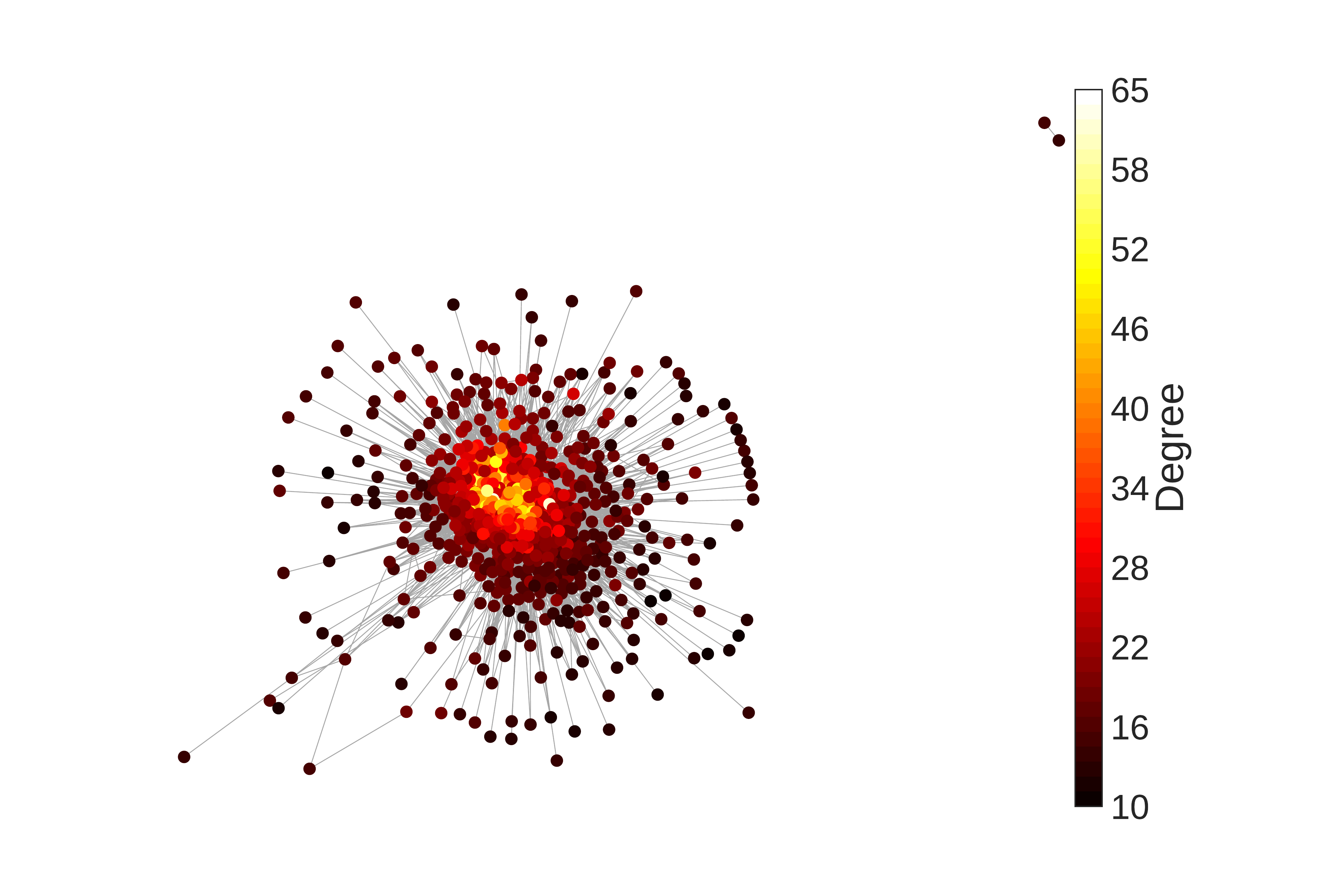}
	\end{subfigure}\\
	\begin{subfigure}[b]{0.5\textwidth}
		\includegraphics[scale=0.26]{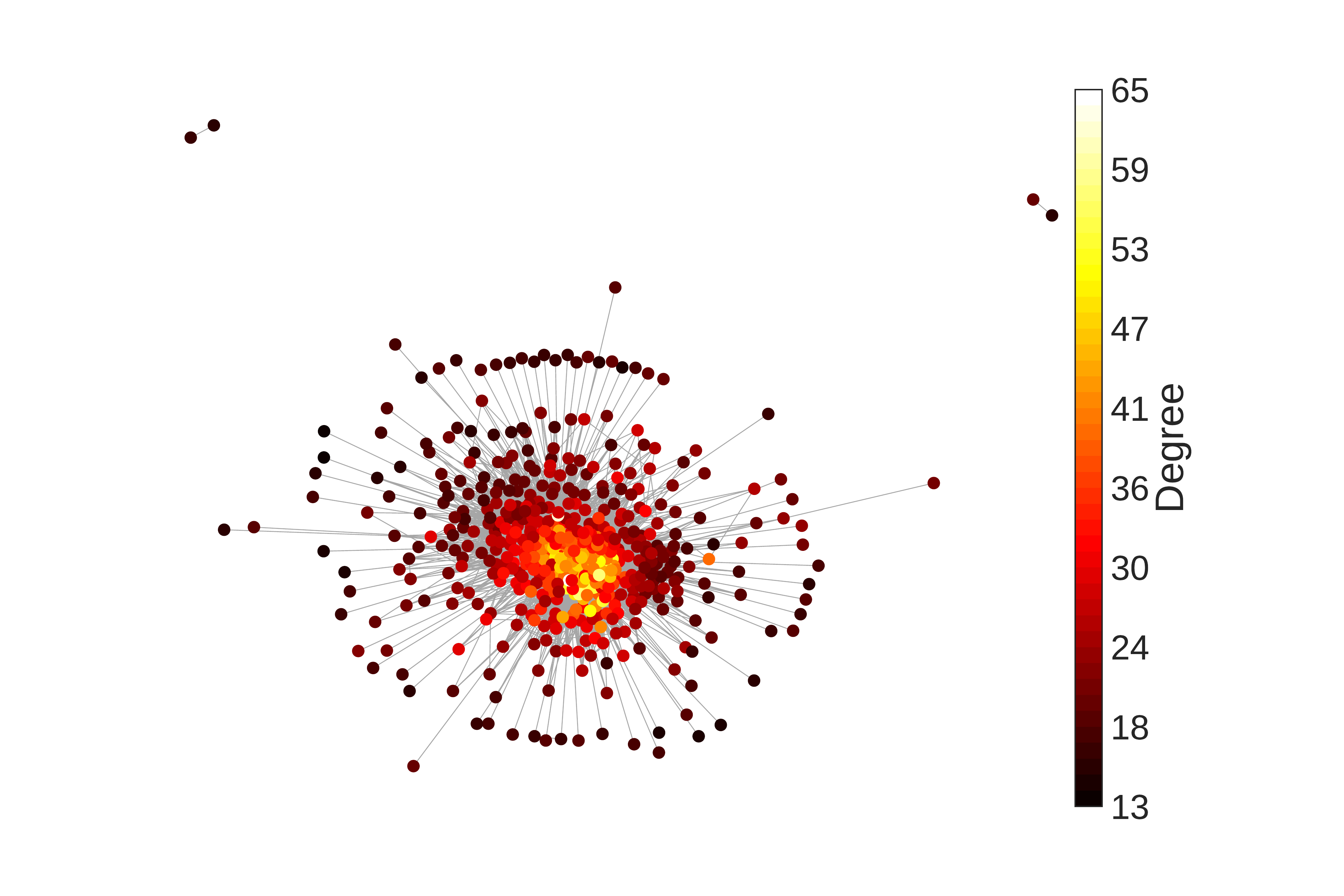} 
	\end{subfigure}
\end{tabular}
	\caption{BiRG (top) and BiPCM$_\text{c}$ (bottom) product networks for the year 2000. The networks are dominated by the largest connected components whose cores are composed of high degree nodes. The degree values refer to the original country-product bipartite network.}
	\label{fig:birg_bipcmc_prod_nws}
\end{figure}
\begin{figure}
\begin{tabular}{c}
	\begin{subfigure}[b]{0.5\textwidth}
		\includegraphics[scale=0.2]{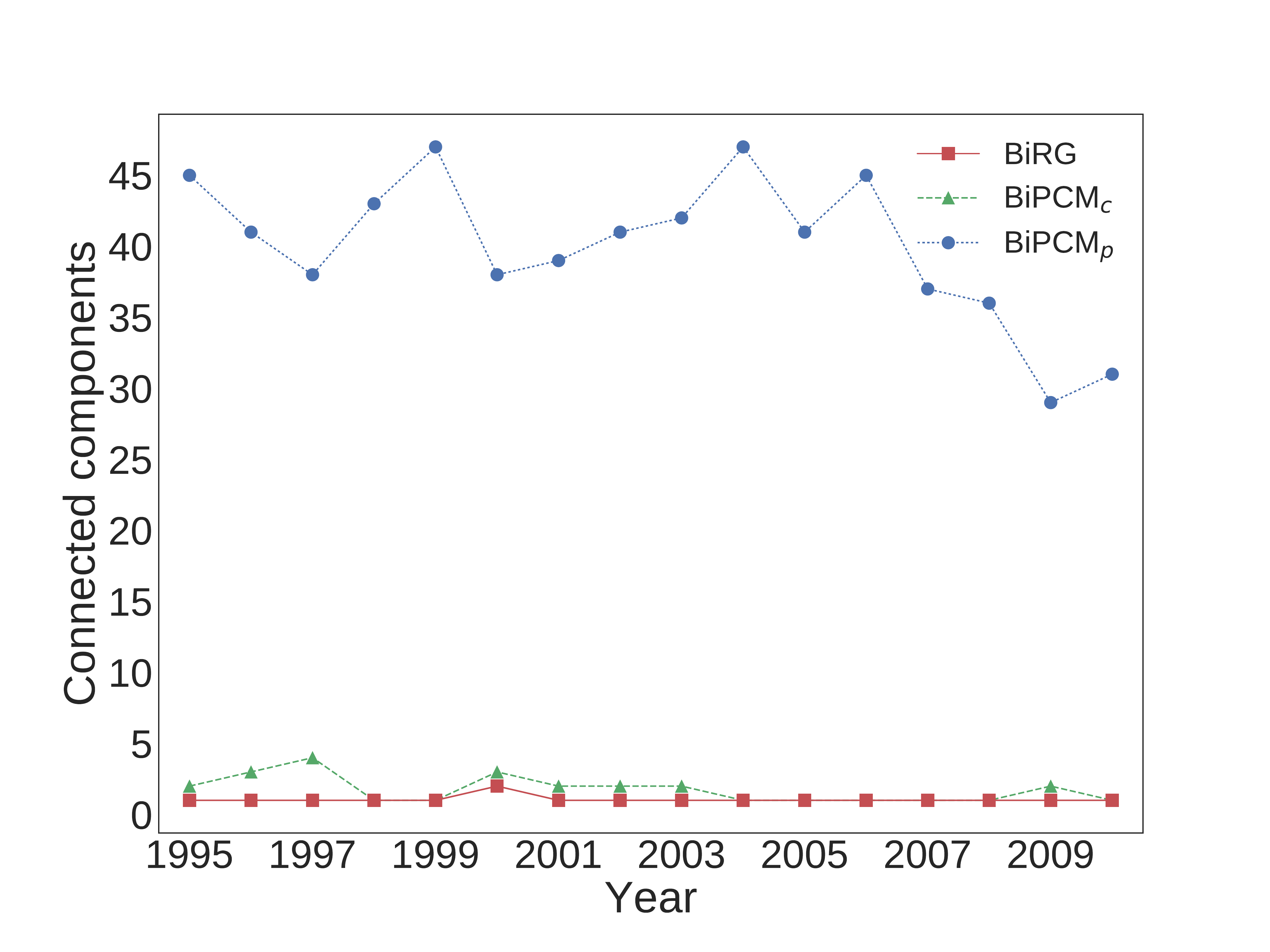}
	\end{subfigure}\\
	\begin{subfigure}[b]{0.5\textwidth}
		\includegraphics[scale=0.2]{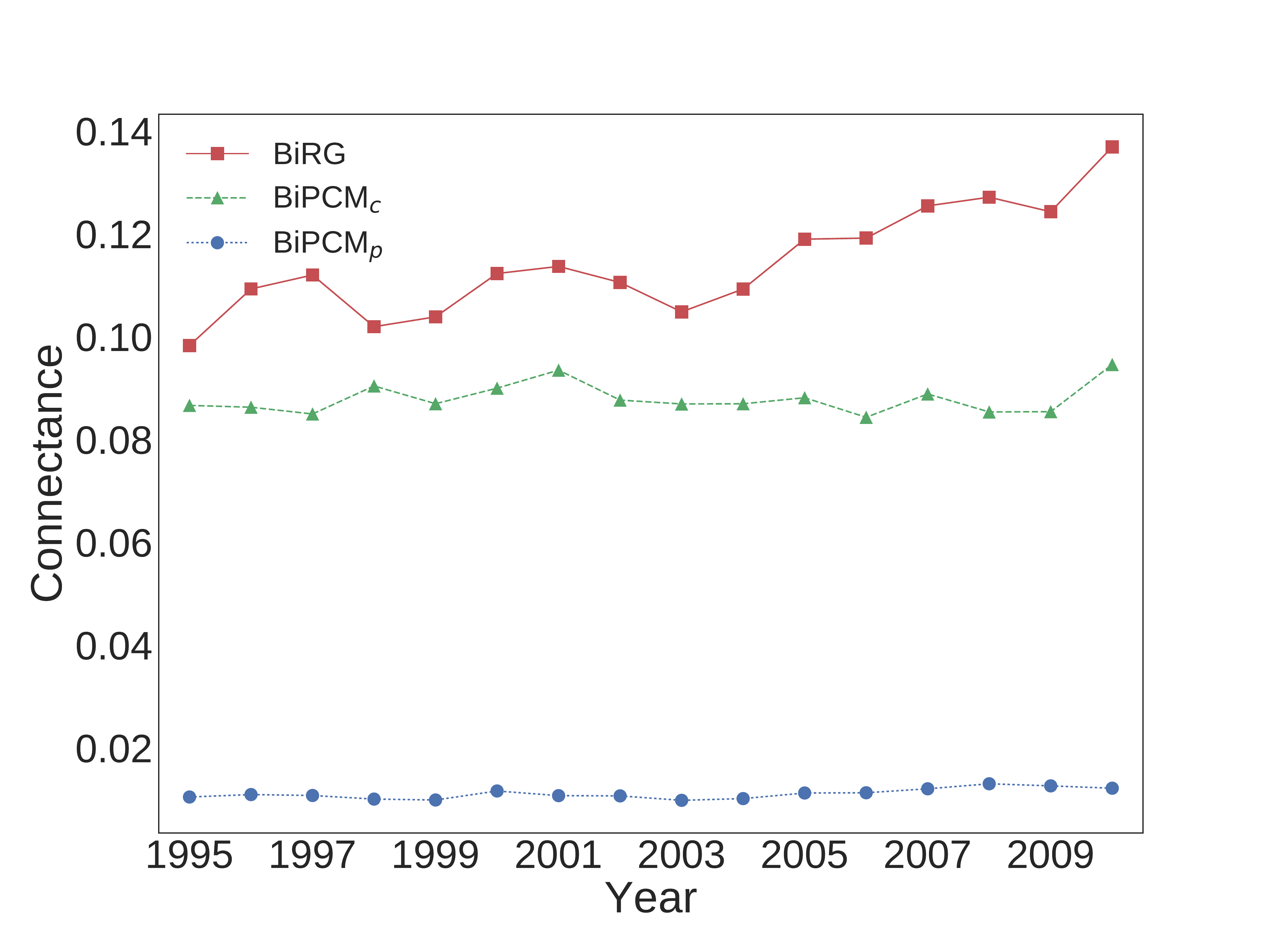}
	\end{subfigure}
\end{tabular} 
	\caption{Properties of the product networks spanned by the
	statistically significant edges according to the respective
	null-models. The BiPCM$_\text{p}$ network is highly fragmented, as
	shown by the comparatively large number of connected components (top)
	and the low connectance (bottom). On the other hand, both BiRG and
	BiPCM$_\text{c}$ are composed of comparatively densely connected
	clusters. Isolated nodes are ignored in both figures.}
	\label{fig:comdet_figs}
\end{figure}

\noindent
The performance of the grand canonical projection algorithm depends on the choice
of the null-model, which defines the information of the original bipartite
network to be discounted in the link verification process. As already mentioned
in the main text, the BiCM imposes the most stringent constraints. For
comparison with the BiPCM$_\text{p}$ product network, Fig.~\ref{fig:birg_bipcmc_prod_nws} illustrates the product
networks obtained if the BiPCM$_\text{c}$ and the BiRG are applied, i.e. if the nodes
of the country layer or the total number of edges are fixed, respectively. It
is easy to see that the two are topologically very different from Fig.~\ref{fig:bicmi_fdr_nw_a001}:
while the BiPCM$_\text{p}$ network is highly fragmented, the BiRG and BiPCM$_\text{c}$ networks are
dominated by the presence of a large connected component, which contains almost all the nodes. The few isolated
clusters are composed of (``meat of swine'', ``pig fat'') and (``cocoa paste'',
``cocoa butter''), and of (``chromium oxides and hydroxides'', ``salts of
oxometallic or peroxometallic acids''), respectively. These product couples are thus
extraordinarily often exported together compared to others. 
The difference between the models is also shown in Fig.~\ref{fig:comdet_figs}.
While the BiRG acts as a relatively ``coarse'' filter, the statistical
verification becomes more strict passing from the BiPCM$_\text{c}$ to the BiPCM$_\text{p}$
and ultimately to the BiCM, for which no links are verified. This observation
is substantially due to the fact that the node-specific probability
distributions of the $V^{ij}$-motifs collapse into a single distribution for
the BiRG and the BiPCM$_\text{c}$, which turn out to be Binomial and Poisson
Binomial~\cite{Saracco2016}. Consequently, the null-models induce a one-to-one
mapping of the $V^{ij}$ measurements onto the \emph{p}-values. Imposing a
significance level for hypothesis testing amounts therefore to choosing a
threshold value $V_{ij, th}$ and discarding motifs with $V_{ij} <
V_{ij, th}$. For the BiRG, $V_{ij, th}\in \{9, 10\}$, whereas for the
BiPCM$_\text{c}$  $V_{ij, th}\in \{12, 13, 14\}$, depending on the year
in the interval 1995 - 2010. As a consequence, only products with $V_{ij}
\geq V_{ij, th}$ bear significant similarity. The only difference between
the motif validations with BiRG and BiPCM$_{\text{c}}$ is a shift in the
\emph{p}-value threshold. The cores of the projection networks host almost exclusively nodes with degrees in the original bipartite network, as one can
confirm by closer inspection of Fig.~\ref{fig:birg_bipcmc_prod_nws}.
It is worth pointing out that several edges in
the BiRG model have \emph{p}-values which are smaller than the machine
precision $\simeq 2.22\cdot10^{-16}$.\\
\end{document}